\documentclass[%
aip,cha,
amsmath,amssymb,
reprint,%
]{revtex4-2} 

\usepackage{graphicx}
\usepackage{bm}
\usepackage[usenames]{color}
\usepackage[utf8]{inputenc}
\usepackage[T1]{fontenc}
\usepackage{mathptmx}
\usepackage{listings}
\usepackage{parskip}
\usepackage{hyperref}
\usepackage{amsmath}  
\usepackage{amsfonts} 
\usepackage[usenames]{color}
\usepackage{soul} 
\usepackage{ulem}

\usepackage{setspace} 
\usepackage[dvipsnames]{xcolor} 

\begin{document}
\renewcommand{\thefootnote}{\alph{footnote}}

\newcommand{\eg}{\textit{e.g.}{~}}
\newcommand{\ie}{i.e.{~}}
\newcommand{\vphi}{\varphi}
\newcommand{\bq}{\begin{equation}}
\newcommand{\ba}{\begin{eqnarray}}
\newcommand{\eq}{\end{equation}}
\newcommand{\ee}{\end{equation}}
\newcommand{\ea}{\end{eqnarray}}
\newcommand{\tchi} {{\tilde \chi}}
\newcommand{\tA} {{\tilde A}}
\newcommand{\sech} { {\rm sech}}
\newcommand{\pstar}{\mbox{$\psi^{\ast}$}}
\newcommand {\bPsi}{{\bar \Psi}}
\newcommand {\bpsi}{{\bar \psi}}
\newcommand {\tu} {{\tilde u}}
\newcommand {\tv} {{\tilde v}}
\newcommand{\dq}{{\dot q}}

\title{}
\title[Stability of nonlinear Dirac solitons under the action of external potential]{Stability of nonlinear Dirac solitons under the action of external potential}

\author{David Mellado-Alcedo}
\email[Electronic mail: ]{alcedo@ugr.es}
\affiliation{Instituto Carlos I de F\'{\i}sica Te\'orica y
	Computacional and Departamento de F\'{\i}sica At\'omica,
	Molecular y Nuclear, Facultad de Ciencias, Universidad de Granada, Av. Fuente Nueva s/n, 18071 Granada, Spain}
	\affiliation{Departamento de F\'{\i}sica Aplicada, Facultad de Ciencias, Universidad de Cordoba, Campus Rabanales, 14071 C\'ordoba, Spain}
 \author{Niurka R.\ Quintero}
 \email[Electronic mail: ]{niurka@us.es} 
 \affiliation{F\'\i sica Aplicada I, Escuela Técnica Superior de Ingeniería Informática, Universidad de Sevilla, 
 	Av. Reina Mercedes s/n, 41012 Sevilla, Spain}

\date{\today}

\begin{abstract}
The instabilities observed in direct numerical simulations of the Gross-Neveu equation under linear and harmonic potentials are studied.
The Lakoba algorithm, based on the method of characteristics, is performed to numerically obtain the two spinor components. 
	We identify non-conservation of energy and charge in simulations with instabilities and 
	we find that all studied solitons are numerically stable, except the low-frequency  solitons oscillating in the harmonic potential over long periods of time. These instabilities, as in the case of Gross-Neveu equation without potential, can be removed  by imposing  absorbing boundary conditions. 
The dynamics of the  soliton is in perfect agreement with the prediction obtained using an ansatz with only two collective coordinates, namely the position and momentum of the center of mass.  We use the temporal variation of both field energy and momentum to determine the evolution equations satisfied by the collective coordinates. By applying the same methodology, we also demonstrate the spurious character of the reported instabilities in the Alexeeva-Barashenkov-Saxena model under external potentials.
\end{abstract}

\date{ \today}

\maketitle

\begin{quotation}
The nonlinear Dirac equation in $1+1$-dimensions supports localized  
solitons. Theoretically, these traveling waves propagate  
with constant velocity, energy, momentum, and charge. However, 
the soliton profiles can be distorted, and eventually destroyed, due to  intrinsic and/or numerical instabilities.  
The constants of motion and the initial profiles can also be modified by external potentials, that may give  rise, in some cases, to  instabilities. 
 The perturbations can render the soliton unstable (intrinsic instability). The difference between initial condition and exact solution of the  perturbed nonlinear Dirac equation, the boundary conditions, and the relationship between the space and time steps, among other reasons, may create numerical instabilities (spurious instabilities). 
Lakoba's numerical algorithm  shows that solitons of the Gross-Neveu equation (nonlinear Dirac equation with scalar-scalar self-interaction) are stable regardless of their frequency, whereas, by using either  operator splitting or Runge-Kutta methods,  
 instabilities for low frequencies are observed (due to the interaction of the soliton with its own dispersive radiation).  
In this work, Lakoba's method of characteristics is employed to numerically solve the massive Gross-Neveu and the  Alexeeva–Barashenkov–Saxena models under spatial potentials.  
It is shown that, in both systems, the initial soliton  remains stable,  in most of its domains of existence. 
This enables it to be concluded that 
the  reported instabilities in the aforementioned two systems were numerical, and that their solitary waves are stable within the range of  
parameters studied.   
	\end{quotation}
	
\section{Introduction\label{intro}} 
The spectral stability analysis of nonlinear waves starts with the linearized equations describing small deviations around the static or stationary nonlinear wave. The stability of the waves is determined by the sign of the real part of the eigenvalues of the corresponding  Sturm-Liouville problem and  the predicted instability can be observed numerically. This is the standard procedure employed to demonstrate the stability (or instability) of the kinks in nonlinear Klein-Gordon equations \cite{parmentier:1967,scott:1969,rubinstein:1970}, of the enveloped solitons in the nonlinear Schr\"odinger equation \cite{zakharov:1967,vakhitov:1973}, and of the Gross-Neveu (GN) solitons  \cite{chugunova:2007,berkolaiko:2012}, just to mention  few specific examples. Although, in the latter   case,  no known analytical solution of the eigenvalue problem has yet been found, semi-analytical calculations \cite{berkolaiko:2012} suggest that the soliton of the massive Gross-Neveu model is linearly stable regardless of its frequency, $\omega$. These semi-analytical results, however, contradict numerical simulations reported earlier in Ref. \cite{bogolubsky:1979},  where the soliton was observed to be unstable for certain frequencies.  

The detailed and systematic numerical simulations performed  in Ref. \cite{shao:2014}  by using the  
operator splitting method \cite{xu:2013} support the existence of instabilities in the Gross-Neveu equation for certain frequencies.  However, Lakoba has recently shown that these instabilities are spurious,  and interestingly, they can be removed through the application of  a novel numerical algorithm based on the method of characteristics.  Furthermore, for low frequencies, it was necessary to add absorbing boundary conditions (ABC) to completely remove these instabilities (for a detailed description  see Refs.  \cite{lakoba:2018,lakoba:2021}). In these studies, it was shown that the spectral method for computing the spatial derivative in the Gross-Neveu equation \cite{cuevas:2015} also leads to numerical instabilities over long periods of time. 

The issue regarding the stability of  the nonlinear waves under perturbations is even more complex. In fact,  
the first drawback encountered is that, in general, there is no exact solitary wave solution in the Gross-Neveu model  under spatial potentials. Therefore, the standard procedure, employed in the unperturbed system in order to investigate the stability of solitary waves, cannot be applied. Instead, the numerical simulations and the collective coordinate theories can help to shed light on this problem \cite{bogdan:1985,quintero:2019b}.

 In Ref.\cite{nogami:1995}, it was shown that the center of mass of the nonlinear Dirac soliton under linear potential moved as a relativistic particle. Furthermore, a small deviation from the classical point representation was found in the case of the harmonic potential. By means of an ansatz with the position and the momentum of a center of mass as collective coordinates, and from a variational approach, the equations of motion which obey the two collective variables were obtained \cite{nogami:1995}. 
These results were confirmed with numerical simulations over relatively short final integration times and for specific values of the soliton's frequency without reporting any instability. A more exhaustive numerical simulation, however, detected certain discrepancies with the approximated analytical results following the  appearance of instabilities (see Figs. 9, 10, 12, and 13 of  Ref. \cite{mertens:2012}). These unstable solutions were related with the observed instabilities of the unforced nonlinear Dirac equation \cite{bogolubsky:1979,shao:2014}.  Nevertheless, subsequent to the recent studies carried out by Lakoba \cite{lakoba:2018,lakoba:2020}, this claim is  no  longer valid and 
the cause of observed instabilities  in Ref. \cite{mertens:2012} can naturally be questioned. 
For the same reasons, the instabilities observed in the Alexeeva-Barashenkov-Saxena (ABS) solitary waves with the same type of potentials  should be investigated (see Figs. 3 and 5 in Ref. \cite{mertens:2021}). 
 It is worth to mention that the soliton of the ABS model is unstable in the low-frequency limit \cite{alexeeva:2019}, and hence it is expected that it remains unstable under the action of the spatial potentials. 
 
The aim of our study is to investigate the 
  soliton dynamics in the Gross-Neveu and Alexeeva-Barashenkov-Saxena models under both linear and harmonic spatial potentials to ascertain the numerical stability of these nonlinear  waves. 
  	The numerical integration of the differential equations is based in the method of characteristics  \cite{lakoba:2018}. Since the potential term  affects each of the two spinor components of the nonlinear equations,   
  the method of
characteristics developed for the unforced Gross-Neveu model can easily be extended. 
These results are complemented by using the aforementioned ansatz with two collective coordinates, 
although the equations of motion have been obtained by means of the continuity equations of the field momentum and energy.  
 The main advantage of this methodology is that avoids the determination of the Lagrangian density as a function of the spinor components. It has been successfully used in the past in other nonlinear systems   \cite{mclaughlin:1978,karpman:1979,maimistov:1993,quintero:2010} in order to describe the dynamics of solitary waves under perturbations.

The outline of the paper is as follows. In Sec. \ref{sec2}, the Gross-Neveu and Alexeeva-Barashenkov-Saxena equations under spatial potential are reviewed, and an  
analysis of the conservation laws is provided. In Sec. \ref{sec3}, an approximated solution is obtained by the method of moments. A numerical
schema based on Lakoba's method of characteristics is developed in Sec. \ref{sec4}, and the   
simulations  of the solitary waves are discussed in Sec. \ref{sec5}. 
Finally, the main conclusions of our results are summarized in Sec. \ref{sec6}.

\section{GN and ABS equations under the action of potential} \label{sec2}

For simplicity, we start with the GN equation (nonlinear Dirac equation in 1+1 dimension with 
scalar-scalar  self interaction 
$\frac{1}{2} ( {\bPsi} \Psi)^{2}$) under the action of vector potential
\bq
i \gamma^\mu \partial_\mu \Psi -  \Psi +  (\bar \Psi \Psi) \Psi - e \gamma^{\mu} A_{\mu} \Psi = 0,  \label{nldev}
\eq
where $\Psi(x,t)=\{\psi(x,t);\chi(x,t)\}^{\scriptscriptstyle T}$ is a vector field with two spinor components, $\bPsi=(\Psi^{\star})^{\scriptscriptstyle T} \gamma^{0}$ is the Dirac adjoint spinor, 
  $A_1=0$, $eA_0 =   V(x)$ are the axial gauge, and 
\bq \label{q1}
\gamma^{0} = \sigma_{3}= 
\left( \begin{array}{cc}
	1 & 0  \\
	0 & -1  \end{array} \right), \qquad \gamma^{1} = i \sigma_2= 
\left( \begin{array}{cc}
	0 & 1  \\
	-1 & 0  \end{array} \right),
\eq
represent the 1+1 dimensional Dirac Gamma matrices, whereby $\sigma_2$ and $\sigma_3$ are the Pauli matrices  \cite{alvarez:1981,mertens:2012}.  
Furthermore, throughout the text, nonreflecting boundary conditions are assumed, that is, the two components of $\Psi(x,t)$ vanish at the boundaries, i.e. 
\begin{equation}
\lim_{x\rightarrow\pm \infty}\Psi(x,t)=0\text{.}
\end{equation}
The corresponding adjoint  equation of \eqref{nldev} reads
\bq
- i \partial_\mu \bar{\Psi} \gamma^\mu -  \bar{\Psi} +(\bar \Psi \Psi) \bar{\Psi} - e \bar{\Psi}  \gamma^{\mu} A_{\mu} = 0.  \label{anldev}
\eq
The energy-momentum tensor is given by the relationship
\begin{eqnarray}\label{emt}
\nonumber &&	T^{\mu \nu}[\psi(x,t),\chi(x,t)]= \frac{i}{2} \left[ \bPsi \gamma^\mu \partial^ \nu \Psi  -  \partial^\nu \bPsi  \gamma^\mu  \Psi \right]  \\ 
	&-& g^{\mu \nu}
	 \left(\frac{i }{2}\right) [\bPsi \gamma^{\mu} \partial_{\mu} \Psi-\partial_{\mu} \bPsi \gamma^{\mu} \Psi] \\ \nonumber
	&-&g^{\mu \nu}( - \bPsi \Psi  
	+ \frac{1}{2} (\bPsi \Psi)^{2} - e \bPsi \gamma^\mu A_\mu \Psi)  \>,
\end{eqnarray}
where 
\bq \nonumber 
g^{\mu \nu} = 
\left( \begin{array}{cc}
	1 & 0  \\
	0 & -1  \end{array} \right). 
\eq
The notation ${\cal B}[\psi(x,t),\chi(x,t)]$ denotes functional dependence of ${\cal B}$ on the spinor components 
$\psi(x,t)$, and $\chi(x,t)$.
The continuity equations with source terms can be obtained by multiplying Eq. (\ref{nldev}) to the left by $\partial_{\nu}\bar{\Psi}$, and Eq. (\ref{anldev}) to the right by $\partial_{\nu}\Psi$, and then adding  these two  expressions 
\cite{mertens:2012}. 
Specifying $\mu=0$, the values $\nu=1$ and $\nu=0$ yields to the continuity equations for the momentum and for the energy, respectively. 

The continuity equation for the momentum reads 
\begin{eqnarray}
&&\dfrac{\partial T^{01}}{\partial t}[\psi(x,t),\chi(x,t)]+ \dfrac{\partial T^{11}}{\partial x}[\psi(x,t),\chi(x,t)]= \nonumber \\
&=& V(x)\,\dfrac{d}{dx}(|\psi(x,t)|^2+|\chi(x,t)|^2).  \label{c1} 
\end{eqnarray}
The functional  dependence of the field momentum of the two spinor components $\psi(x,t)$ and $\chi(x,t)$ is defined by
\bq \label{eq:momentum}
P[\psi(x,t),\chi(x,t)] =\int_{-\infty}^{+\infty}  dx T^{01}[\psi(x,t),\chi(x,t)]. 
\eq	
By integrating Eq. ({\ref{c1})  across the whole space, and by assuming that the condition 
\bq \label{eq:bc}
\lim_{x \to + \infty}C(x,t)  -  \lim_{x \to - \infty}C(x,t) = 0,
\eq 
is fulfilled for $C(x,t)=T^{11}(x,t)$,  
it is obtained the following time evolution of the momentum 
	\bq \label{eq:conmome}
	\frac {dP}{dt} = - \int_{-\infty}^{+\infty}  dx \dfrac{dV}{dx} \bar{ \Psi} \gamma^{0} \Psi. 
	\eq
	Note that, for the linear potential, this equation can be exactly integrated regardless of the ansatz. 
	Therefore, the evolution of the momentum will be employed to verify numerical results for this particular case.
	
	Since the potential is time-independent, the energy 
	\bq \label{eq:ene}
	E[\psi(x,t),\chi(x,t)] =\int_{-\infty}^{+\infty}  dx T^{00}[\psi(x,t),\chi(x,t)],
	\eq		
	is conserved. Indeed, by integrating the continuity equation for the energy over space 
	\bq
	\dfrac{\partial T^{00}}{\partial t}[\psi(x,t),\chi(x,t)]+ \dfrac{\partial T^{10}}{\partial x}[\psi(x,t),\chi(x,t)]=  0, \label{c2}
	\eq
	and assuming that Eq. \eqref{eq:bc} is satisfied for $C(x,t)=T^{10}(x,t)$, 
the conservation of energy, \ie  $dE/dt=0$, is shown.
	
	Furthermore, by multiplying   Eq. \eqref{nldev} to  the left by $\bar{\Psi}$ and Eq. \eqref{anldev} to the right by $-\Psi$, and subsequently adding them together,  
	the continuity equation for the charge reads
	\bq \label{eqc:charge}
	\dfrac{\partial \rho}{\partial t}(x,t)+\dfrac{\partial j}{\partial x}(x,t) =0, 
	\eq
	being $\rho(x,t)=\bar{\Psi}\gamma^0 \Psi$ the 
charge density and $j(x,t)=\bar{\Psi}\gamma^1 \Psi$  the  current density. Therefore, 
the charge is calculated by performing the following integral 
\bq \label{eq:charge}
Q[\psi(x,t),\chi(x,t)] =\int_{-\infty}^{+\infty}  dx \, \rho(x,t). 
\eq	
By integrating Eq. \eqref{eqc:charge} over $x$, and by assuming that Eq.\ \eqref{eq:bc} is fulfilled for $C(x,t)=j(x,t)$, 
it is shown that the charge 	
is also conserved, \ie, $dQ/dt=0$.  
 The conservation of the charge, as well as the energy, allows to confirm the convergence of numerical results. 
	 
In terms of the two spinor components $\Psi(x,t) = \{\psi(x,t); \chi(x,t)\}^{T}$, Eq.\ \eqref{nldev} is equivalent to  
	\ba
	\dfrac{\partial \psi}{\partial t} + \dfrac{\partial \chi}{\partial x} &=&i\,[|\psi|^2-|\chi|^2 -1 -V(x)] \psi, \label{eqs1} \\ 
	\dfrac{\partial \chi}{\partial t} + \dfrac{\partial \psi}{\partial x} &=&-i\,[|\psi|^2-|\chi|^2 -1 +V(x)] \chi. \label{eqs2}
	\ea
	Following the procedure of Ref. \cite{lakoba:2018}, the change of dependent variables 
	\ba
	u = \dfrac{\psi + \chi}{\sqrt{2}}, \quad v = \dfrac{\psi - \chi}{\sqrt{2}},
	\ea
	is  assumed, and the Eqs.\ \eqref{eqs1}-\eqref{eqs2}  can be rewritten as
	\ba
	\dfrac{\partial u}{\partial t} + \dfrac{\partial u}{\partial x} &=&i\,[u\,v^\star+u^\star\,v -1] v -i\,V(x)\,u, \label{equ} \\ 
\dfrac{\partial v}{\partial t} - \dfrac{\partial v}{\partial x} &=&i\,[u\,v^\star+u^\star\,v -1] u -i\,V(x)\,v. \label{eqv}
	\ea
A modification of the nonlinear term and  the mass term on these equations leads to the ABS model. Indeed, the two spinor components $\{v(x,t),u(x,t)\}^{T}$ of the ABS model \cite{alexeeva:2019} under spatial potential $V(x)$ satisfy the two coupled and nonlinear partial differential equations \cite{mertens:2021} 
	\ba
	\dfrac{\partial u}{\partial t} + \dfrac{\partial u}{\partial x} &=&i\,[u^\star\,v +1] v -i\,V(x)\,u, \label{equc} \\ 
\dfrac{\partial v}{\partial t} - \dfrac{\partial v}{\partial x} &=&i\,[u\,v^\star +1] u -i\,V(x)\,v. \label{eqvc}
\ea	
For $V(x)=0$, it can be exactly solved (see the details in Ref. \cite{alexeeva:2019}).  
It is worth  mentioning that 
 the momentum $P[v(x,t),u(x,t)]$, the energy $E[v(x,t),u(x,t)]$  
(up to a term related with the nonlinearity) and the charge 
$Q[v(x,t),u(x,t)]$, of the ABS model have the same  functional form defined 
by Eqs.\ \eqref{eq:momentum}, \eqref{eq:ene}, and \eqref{eq:charge}, respectively.  Here, the energy density is defined by
\begin{eqnarray} 
	\nonumber 
	&& T^{00}[v(x,t),u(x,t)]=\frac{i}{2}\,\left[u\,\dfrac{\partial u}{\partial x}^\star+v^\star\,\dfrac{\partial v}{\partial x}-u^\star\,\dfrac{\partial u}{\partial x}-v\,\dfrac{\partial v}{\partial x}^\star\right]\\ 
	&&
	-(u\,v^\star+u^\star\,v)-\frac{(u\,v^\star)^2+(u^\star\,v)^2}{2}+V(x) [|u|^2+|v|^2]. \label{eq7b} 
\end{eqnarray}
As in the GN model, the momentum $P[v(x,t),u(x,t)]$ satisfies  Eq. \eqref{eq:conmome}, while the energy $E[v(x,t),u(x,t)]$ and the charge $Q[v(x,t),u(x,t)]$ are also conserved. 
These common features in the dynamics between the  two nonlinear Dirac equations resemble the similar behavior of the kinks in the nonlinear Klein-Gordon equations with and without perturbations, where the collective variables of sine-Gordon, the double sine-Gordon, and the $\varphi^4$ kinks share the same equations of motion with different coefficients \cite{salerno:2002,quintero:2001,morales-molina:2006}. 
 In the next section, it is shown that this is exactly what happens when the equations of motion for the collective coordinates for GN and for ABS models are compared. 

\section{Method of moments} \label{sec3} 
   
Let us focus on the GN model, represented by Eqs.\ \eqref{equ}-\eqref{eqv}. 
 We start from 
the exact moving soliton of Eq.\ \eqref{nldev} with $A_{\mu}=0$, which is represented by 
\ba
&&
\psi(x,t) = \left( \cosh{\frac{\alpha}{2}} A(x') 
+ i \sinh{\frac{\alpha}{2}} B(x') \right) e^{-i \omega\,t'} , \label{eq:psi} \\
&&\chi(x,t) = \left( \sinh{\frac{\alpha}{2}} A(x') 
+ i \cosh{\frac{\alpha}{2}} B(x') \right) e^{-i \omega\,t'} , \label{eq:chi}
\ea
where $x' = \gamma~(x-v_s\, t)$ and $t'= \gamma (t - v_s\,x)$ are the Lorentz transformation, 
$v_s$ is the soliton velocity,  and $\gamma=\cosh(\alpha)=1/\sqrt{1-v_s^2}$ is the well-known Lorentz factor \cite{lee:1975}. 
Moreover, 
\begin{eqnarray}
	\label{eqAc}
	A(x)&=& \sqrt{2} \beta \frac{\sqrt{1+\omega} \cosh(\beta x)}
	{1 + \omega \cosh(2 \beta x)}, \\
	\label{eqBc}
	B(x)&=& \sqrt{2} \beta \frac{\sqrt{1-\omega} \sinh(\beta x)}
	{1 + \omega \cosh(2 \beta x)},  
\end{eqnarray} 
with $\beta=\sqrt{1-\omega^2}$ and the frequency $\omega \in(0,1)$. 

	When the potential $V(x)$ is added to the nonlinear Dirac equation, an ansatz is assumed as an approximated solution which is obtained through a slight modification of the exact solution.  
 Specifically, for the 
GN equation under the spatial potential \eqref{nldev},        
the trial wave function in component form is given by Eqs.\ \eqref{eq:psi}-\eqref{eq:chi}, 	
where $x'$ is replaced with $z= \gamma(t)~(x-q(t))$, $\gamma(t)=1/\sqrt{1-\dot{q}^2(t)}$ denoting by $\dot{q}=dq/dt$, and 
$\omega\,t'$ is replaced with $\omega\,\gamma(t) t + p(t)\,x$. 
In principle, $\omega$ could be replaced by $\omega(t)$, however, the conservation of the charge implies that $\omega$ is time-independent. 
The two collective coordinates used here are the position $q(t)$ and the momentum $p(t)$ of the center of mass. Therefore, the ansatz is represented by
\ba
&&
\tilde{\psi}(z,t) = \left(\cosh{\frac{\alpha(t)}{2}} A(z) 
+ i \sinh{\frac{\alpha(t)}{2}} B(z) \right) e^{i\,\phi(z,t)}, \label{eq:psia} \\
&&\tilde{\chi}(z,t) = \left(\sinh{\frac{\alpha(t)}{2}} A(z) 
+ i \cosh{\frac{\alpha(t)}{2}} B(z) \right) e^{i\,\phi(z,t)}, \label{eq:chia}
\ea
where $\phi(z,t)=p(t)[z/\gamma(t)+q(t)]-\omega\gamma(t)t$, $\cosh \alpha(t)=\gamma(t)$, and here  
$q(t)$ and $p(t)$ are unknown functions.

The one collective coordinate 
ansatz used in Ref. \cite{nogami:1995}, that replaced $\omega\,t'$ with $\omega\,\gamma(t) (t - \dot{q}(t)\,x)$, is inconsistent because  the solution obtained from the evolution equation of the momentum fails to fulfill the conservation of energy unless the potential is linear.

By inserting the ansatz in $T^{00}$ given by  \eqref{emt} and integrating over $x$, the energy
\bq \label{eq:ea}
\tilde{E} =M_0\,\gamma(t)+U(q,\dot{q})+Q\,\dot{q}\,(p-\omega\,\dot{q}\,\gamma)
\eq  
is obtained as functions of the collective  coordinates. 
 The charge, $Q$, and the mass at rest,
$M_0$, are  constants determined by the frequency $\omega$ (see Table \ref{tab1}).
\begin{table} [h!]
	\begin{tabular}{|c|c|c|}
		\hline
		Magnitudes   &     Gross-Neveu  &  ABS \\
		\hline
		$\omega$  &  $0<\omega<1$ & $1/\sqrt{2}<\omega<1$ \\		
		\hline
		$Q$ & $\dfrac{2\,\beta}{\omega}$& $2\,\ln\left[\dfrac{\omega+\beta}{\omega-\beta}\right]$\\
		\hline
		$M_0$ & $4\, \tanh^{-1}\left(\sqrt{\dfrac{1-\omega}{1+\omega}}\right)$ & $\sqrt{2}\ln\left[\dfrac{1+\sqrt{2}\beta}{1-\sqrt{2}\beta}\right]$ \\
		\hline
	\end{tabular}
	\caption{Values of  the frequency $\omega$,  charge $Q$, and mass at rest $M_0$ for the GN and ABS models without perturbations.}
	\label{tab1}
\end{table}
Moreover, the particle potential  
\bq \label{eq:pot}
U(q,\dot{q})= \int_{-\infty}^{+\infty} dz\,\dfrac{\rho(z,t)}{\gamma(t)}\, V\left( \dfrac{z}{\gamma}+q(t)\right),
\eq
is obtained by weighting the external potential with the density of the charge $\rho(z,t)=\gamma(t)\,[A^2(z)+B^2(z)]$. 
The evolution of the momentum \eqref{eq:conmome} can be rewritten as
\bq \label{eq:Pat}
\dfrac {d\tilde{P}}{dt} =
-   \dfrac {\partial U}{\partial q}  ,
\eq
where $\tilde{P}(t)$ satisfies
\bq \label{eq:Pa}
\tilde{P}(t)=M_0\,\gamma(t)\,\dot{q}(t)+Q\,(p-\omega\,\dot{q}\,\gamma).
\eq
In terms of the momentum $\tilde{P}(t)$, the energy \eqref{eq:ea} becomes 
\bq \label{eq:ea1}
\tilde{E}(t) =\dot{q}\, \tilde{P}+\dfrac{M_0}{\gamma(t)}+U(q,\dot{q}). 
\eq	
By setting the time derivative of this energy to zero,  
and by substituting $\tilde{P}(t)$ from \eqref{eq:Pa} in the obtained expression, the collective coordinate $p(t)$ is determined by     
\bq \label{eq:p}
p(t)=\omega\,\dot{q}\,\gamma-\dfrac{1}{Q}\dfrac {\partial U}{\partial \dot{q}}.
\eq 
From Eqs.\ \eqref{eq:Pat},  \eqref{eq:Pa}, and \eqref{eq:p}, the evolution of the center of mass of the solitons is governed by the second Newton law 
\bq \label{eq:newton}
\dfrac {d}{dt}[M_0\,\gamma(t)\,\dot{q}(t)]  = \dfrac {d }{dt}\dfrac {\partial U}{\partial \dot{q}}-\dfrac {\partial U}{\partial q}  ,
\eq
in agreement with Eq.\ (5.23) of Ref.\ \cite{mertens:2012}. 

The absence of any perturbation in the GN equation ($V(x)=0$) implies $U=0$, $p=\omega\,\dot{q}\,\gamma$, $\dot{q}=\dot{q}(0)$, where $\dot{q}(0)$  
is the initial constant velocity, and the exact solution, Eqs.\  \eqref{eq:psi}-\eqref{eq:chi}, is recovered. This method presents two advantages over that developed  in Ref. \cite{mertens:2012}. 
First, it can be applied to  systems with unknown Lagrangian density, and second, 
the equations of motion guarantee fulfillment of both continuity equations and conservation laws. 

 This methodology can be extended to the ABS model under spatial potential, 
 Eqs.\ \eqref{equc}-\eqref{eqvc}.  
As the approximated solution of Eqs.\ \eqref{equc}-\eqref{eqvc}, the following ansatz  is  assumed with two collective coordinates, the position $q$ and the momentum $p$, 
\begin{eqnarray}
	\tilde{u}(z,t) &=& -e^{\alpha/2}\,a(z)\,e^{-i\,\theta(z)}\,e^{i\,\phi(z,t)}, \label{eq26a} \\
	\tilde{v}(z,t) &=& e^{-\alpha/2}\,a(z)\,e^{i\,\theta(z)}\,e^{i\,\phi(z,t)}, \label{eq26b}
\end{eqnarray}
where 
\begin{eqnarray}
	\nonumber		
		a^2(z)&=&  
		\frac{[2(1-\omega) \sech^2(\beta z)][1+\lambda^2\,\tanh^2(\beta\,z)]}{1-6 \lambda^2\,\tanh^2(\beta\,z)+\lambda^4\,\tanh^4(\beta z)},     
		\\
	\theta(z)	&=&-\arctan[\lambda \,\tanh(\beta\,z)], 
		\nonumber 
	\end{eqnarray} 
$\lambda=\sqrt{(1-\omega)/(1+\omega)}$, $1/\sqrt{2}<\omega<1$,  and $\alpha$, 
$\beta$, $z$, and $\phi(z,t)$ are defined as in Eqs.\ \eqref{eq:psia}-\eqref{eq:chia}.  By inserting this ansatz in Eqs.\ 
\eqref{eq:momentum} and \eqref{eq7b}, and integrating  
	over $x$, it is obtained that the momentum and the energy are also given by Eqs.\ \eqref{eq:Pa} and \eqref{eq:ea1}, respectively.   
From the time variation of both field momentum and energy, it is straightforward to show that all the equations \eqref{eq:ea}-\eqref{eq:newton} are satisfied, where the density of the charge in Eq.\ \eqref{eq:pot} is given by $\rho(z,t)=|\tilde{u}(z,t)|^2+|\tilde{v}(z,t)|^2=2\,\gamma(t)\,a^2(z)$, and  the new parameters $\omega$, $Q$, and  $M_0$ are collected in  Table \ref{tab1} . The exact solitary wave solution of the ABS model  is recovered from the ansatz by solving the equations  \eqref{eq:ea}-\eqref{eq:newton} with $V(x)=0$. 		

\subsection{Linear potential}
The linear potential $V(x)=-V_1\,x$ in the new variable $z$ reads $V(z)=-V_1\,\left(\dfrac{z}{\gamma}+q\right)$. By inserting  $V(z)$  in Eq.\ \eqref{eq:pot} and by integrating this equation, the potential $U(q)=-V_1\,Q\,q$ is obtained. The particle potential is also linear in $q$, and its slope depends on $Q$. Since $U$ is independent of $\dot{q}$,  from Eq.\ \eqref{eq:p}, $p(t)=\omega\,\gamma\,\dot{q}$. Therefore, the 
two collective coordinates are no longer independent, and the two considered ans\"atze 
become one collective coordinate ansatz. 
By setting $\dot{q}(0)=0$ as initial conditions, the momentum  $\tilde{P}(t)=P(t)=M_0\,\gamma(t)\,\dot{q}(t)$ vanishes at zero, and Eq. \eqref{eq:Pat} becomes  
\bq
\label{eq:cc1}
\dfrac{dP}{dt}=V_1\,Q.  
\eq
Therefore, the momentum reads  
	\bq \label{eq:solp}
P(t)= V_1\,Q\,t,
\eq
which is a linear function of time with positive (negative) slope if $V_1>0$ ($V_1<0$). Consequently, 
from Eqs.\ \eqref{eq:Pa} and \eqref{eq:solp}, the velocity 
	\bq \label{eq:solqdot}
	\dot{q}(t)= \dfrac{V_1\,Q\,t}{\sqrt{M_0^2+(V_1\,Q\,t)^2}} 
	\eq
	is derived. By integrating the velocity, the position of the center of mass reads 
	\bq \label{eq:solq}
	q(t)= q(0)+\dfrac{\sqrt{M_0^2+(V_1\,Q\,t)^2}-M_0}{V_1\,Q}.
	\eq
These results are in agreement with those obtained ones in Refs.\ \cite{nogami:1995,mertens:2012} by using 
a slightly modified ansatz and by deriving the equations of motion by a different procedure. 

\subsection{Harmonic potential}

The harmonic potential $V(x)=(V_2/2)\,x^2$ ($V_2>0$) in the new variable $z$ reads $V(z)=(V_2/2)\,\left(\dfrac{z}{\gamma}+q\right)^2$. 
By inserting  $V(z)$  in Eq.\ \eqref{eq:pot} and by integrating this equation, the particle potential 
 depends not only on the position but also on the velocity, and  has the form
\bq \label{eq:U}
U(q,\dot{q})= \dfrac{V_2}{2}\,\left[q^2\,Q+Q_{2}(1-\dot{q}^2)\right],
\eq
being $Q_2=\int_{-\infty}^{+\infty} dz\, z^2 [A(z)^2+B(z)^2]$  the second moment of the charge. From Eqs.\ \eqref{eq:p} and \eqref{eq:U}, the momentum $p(t)$ reads
\bq \label{eq:p1}
p(t)=\dot{q}(t)\,\left(\omega\,\gamma(t)+V_2\,\dfrac{Q_2}{Q}\right).
\eq
Furthermore, from Eq.\ \eqref{eq:newton}, the position $q(t)$ satisfies
\bq \label{eq:sp}
(M_0\,\gamma^3(t)+V_2\,Q_2)\,\ddot{q}(t)+V_2\,Q\,q(t)=0.
\eq
In the non-relativistic limit, that is, $\dot{q} \ll 1$ (implying $\gamma \approx 1$), Eq.\ \eqref{eq:sp}  becomes  a simple pendulum equation, which has an exact solution (for detailed calculations,  see Ref.\  \cite{mertens:2012}). Clearly, since the center of mass of the soliton is in the harmonic potential, it oscillates around a certain position depending on the initial conditions. We recall that under the assumption $p(t)=\omega\,\gamma\,\dot{q}$  of an ansatz with one collective coordinate \cite{nogami:1995}, the position $q(t)$ is determined, which does not preserve energy $\tilde{E}$. Thus, the simplest ansatz is the one given by Eqs.\ 
\eqref{eq:psia}-\eqref{eq:chia} for the GN model and by Eqs.\ 
\eqref{eq26a}-\eqref{eq26b} for the ABS system.     

\section{Numerical algorithm} \label{sec4}
	Let us unify the two systems of equations \eqref{equ}-\eqref{eqv} and \eqref{equc}-\eqref{eqvc} into a new system of two equations, 
\ba
\dfrac{\partial u}{\partial t} + \dfrac{\partial u}{\partial x} &=&i\,[\kappa \,u\,v^\star+u^\star\,v -m] v -i\,V(x)\,u, \label{equcc} \\ 
\dfrac{\partial v}{\partial t} - \dfrac{\partial v}{\partial x} &=&i\,[u\,v^\star+\kappa\,u^\star\,v -m] u -i\,V(x)\,v, \label{eqvcc}
\ea
where the parameter 
$\kappa$  can take either the value $0$ or $1$ and the parameter $m$ can be either $-1$ or $1$. The GN model is recovered for $\kappa=1$ and $m=1$, while the ABS model is retrieved for $\kappa=0$ and $m=-1$.

It is straightforward to show that 	Eqs.\  \eqref{equcc}-\eqref{eqvcc} are equivalent to 
\ba
\dfrac{\partial u}{\partial \eta} &=&i\,[\kappa \, u\,v^\star+u^\star\,v -m] v -i\,V(\eta-\xi)\,u, \label{equc1} \\ 
\dfrac{\partial v}{\partial \xi}
 &=&i\,[u\,v^\star+\kappa \,u^\star\,v -m] u -i\,V(\eta-\xi)\,v, \label{eqvc1}
\ea
where 
\ba \label{eq:cha} 
\xi=\dfrac{t-x}{2}, \qquad \eta=\dfrac{t+x}{2},
\ea
represent the characteristic coordinates. By setting $m=1$, $\kappa=1$, and $V(x)=0$,  the equations (8b) of Ref. \cite{lakoba:2018} are recovered. 

Although the method of characteristics reduces the partial differential equations \eqref{equcc}-\eqref{eqvcc} 
to the system \eqref{equc1}-\eqref{eqvc1}  along the characteristics \eqref{eq:cha},  
 no exact solutions 
have been found for the linear and harmonic potentials. To proceed,  these equations are  numerically solved.   

Here, we sketch the main steps of the numerical schema proposed by Lakoba  in Ref. \cite{lakoba:2018}.  
The Improved Euler Method is applied to solve the nonlinear Dirac equations under 
 spatial potentials. For simplicity, Eqs.\ \eqref{equcc}-\eqref{eqvcc} are rewritten as 
\begin{equation} \label{eq:sys}
	\left\lbrace
	\begin{array}{l}
		\frac{\partial u}{\partial \eta}=F(\kappa,\eta,\xi,u,v)\\
		\frac{\partial v}{\partial \xi}=F(\kappa,\eta,\xi,v,u)
	\end{array}
	\right. \quad \text{,}
\end{equation}
with  
 $F(\kappa,\eta,\xi,b,c)=i\,[\kappa\,b\,c^\star+b^\star\,c-m]c -i\,V(\eta-\xi)\,b$. The potential 
$V(x)$ enters in the last term of the function $F$. 
This is the reason why the variables $\eta$ and $\xi$ explicitly  appear in the function $F$. Notice that, 
this methodology  can also be applied  
if the potential $V$  depends on $t$. 

Instead of the original infinite domain, the spatial coordinate $x \in [-L,L]$ is bounded. The length of the domain is taken such that $L$ 
is much greater than the soliton width and so that the soliton is far away from the boundaries. The variable $x$ is no longer continuous and takes discrete values defined at the   $k$-spatial-node by $x_{k}$, with $k=1, 2, \dots, K$  being $K=2L/\Delta x+1$ the spatial nodes number. The space step is $\Delta x=x_{k+1}-x_{k}$, and  the spatial nodes at the boundaries  are $x_{1}=-L$ and $x_{K}=L$. 
The time is also discretized  and defined at the n-temporal-node by $t_n$, where $n=1,2, \dots, N$, being $N=t_{N}/\Delta t+1$ the temporal nodes number. The temporal node goes from $t_1=0$  at the initial time until  $t_{N}$ at the 
final  time of integration, 
 being the time step $\Delta t=t_{n+1}-t_{n}$. 

By fixing the point $(x_k,t_n)$ and moving one step in the $x$-direction and one step in the $t$-direction, \ie along the characteristic line where $\bar{\xi} \equiv \xi_{k}^{n}$ is constant,  the point 
$(x_{k+1},t_{n+1})$ is reached.  Likewise, but by moving backward one step in the x-direction and forward one step in the t-direction, \ie along the characteristic line where $\bar{\eta}=\eta^n_k$ is constant, the point $(x_{k-1},t_{n+1})$ is attained. In the following, the notation  $u_{k}^{n}=u(\xi_{k}^{n},\eta_{k}^{n})$ and $v_{k}^{n}=v(\xi_{k}^{n},\eta_{k}^{n})$ is used, with  
\begin{equation}
	\xi_{k}^{n}=\frac{t_{n}-x_k}{2}\text{,} \qquad  \eta_{k}^{n}=\frac{t_n+x_{k}}{2} \text{.}
\end{equation}
 Furthermore, the step sizes in time and space are taken to be equal $\Delta t=\Delta x=h$, which implies  that they are also equal  that the step sizes across the $\eta$ and $\xi$ variables, that is, $\Delta\eta\equiv\eta_{k+1}^{n+1}-\eta_{k}^{n}=h$, and  $\Delta\xi\equiv\xi_{k-1}^{n+1}-\xi_{k}^{n}=h$.
 
Following the algorithm suggested by Lakoba \cite{lakoba:2018}, a predictor-corrector method \cite{birkhoff:1989} is used. 
The first step demands the solution, by using the Simple Euler Method, of the first (second) equation of \eqref{eq:sys} along the characteristic line $\bar{\xi}=\xi_{k}^{n}$ ($\bar{\eta}=\eta_{k}^{n}$), thereby obtaining 
\begin{subequations}
	\begin{align} \label{eq:n1}
		\bar{u}_{k+1}^{n+1}&=u_{k}^{n}+\Delta\eta  F\big(\kappa,\eta_{k}^{n},\bar{\xi},u_{k}^{n},v_{k}^{n}\big) \text{,}\\ \label{eq:n2} 
		\bar{v}_{k-1}^{n+1}&=v_{k}^{n}+\Delta\xi  F\big(\kappa,\bar{\eta},\xi_{k}^{n},v_{k}^{n},u_{k}^{n}\big) \text{,}
\end{align}
\label{equi}
\end{subequations}
where $\bar{u}_{k+1}^{n+1}$ and $\bar{v}_{k-1}^{n+1}$ are the \textit{predicted} values of $u$ and $v$, respectively. The index $k$ takes the values $k=1, 2, \cdots, K-2$ in  Eq.\ \eqref{eq:n1} and the values $k=3, \cdots, K$ in Eq. \eqref{eq:n2}. In both equations, the super-index $n$ takes the values $n=1, 2, \cdots, N-1$.

In a second step, the values of $u$ and $v$ corrected by the Trapezoidal Rules \cite{birkhoff:1989}, ${u}_{k+1}^{n+1}$ and ${v}_{k-1}^{n+1}$, respectively, are obtained:
\begin{subequations}
	\begin{align} \label{eq:uf}
	u_{k+1}^{n+1}&=u_{k}^{n}+\frac{h}{2} 
	\big[F\big(\kappa,\eta_{k}^{n},\bar{\xi},u_{k}^{n},v_{k}^{n}\big)+
	F\big(\kappa,\eta_{k+1}^{n+1},\bar{\xi},\bar{u}_{k+1}^{n+1},\bar{v}_{k+1}^{n+1}\big)\big]
	\text{,} \\ \label{eq:vf}
		v_{k-1}^{n+1}&=v_{k}^{n}+\frac{h}{2} 
		\big[F\big(\kappa,\bar{\eta},\xi_{k}^{n},v_{k}^{n},u_{k}^{n}\big)+
	F\big(\kappa,\bar{\eta},\xi_{k-1}^{n+1},\bar{v}_{k-1}^{n+1},\bar{u}_{k-1}^{n+1}\big)\big]
		 \text{.} 
	\end{align}
	\label{eqvj}
\end{subequations}
Notice that Eq.\ \eqref{eq:uf} connects node $(x_{k},t_n)$ with $(x_{k+1},t_{n+1})$, and Eq.\ \eqref{eq:vf} connects node $(x_{k},t_n)$ with $(x_{k-1},t_{n+1})$, see  Fig.\ \ref{fig:dia}.
   
In order to solve the numerical schema, it is necessary to complement it by adding the initial and boundary conditions fulfilled by $u$ and $v$. The nonreflecting boundary conditions for the finite size of the domain are 	given by $\psi(\pm L,t)=0$ and $\chi(\pm L,t)=0$ (GN  model) and by $u(\pm L,t)=0$ and $v(\pm L,t)=0$ (ABS model), which imply that $u$ and $v$ satisfy the following boundary conditions
\begin{subequations}
	\begin{align}
		u_1^n&=u_K^n=0 \text{,} \\
		v_1^n&=v_K^n=0 \text{.}
	\end{align}
\end{subequations}
 The initial conditions for the Gross-Neveu model are given by 
	\begin{subequations} \label{eq:ic}
		\begin{align}
			u_k^1&=\frac1{\sqrt{2}}\left[\tilde{\psi}(x_k,t_1)+\tilde{\chi}(x_k,t_1)\right] \text{,} \\
			v_k^1&=\frac1{\sqrt{2}}\left[\tilde{\psi}(x_k,t_1)-\tilde{\chi}(x_k,t_1)\right] \text{,}
		\end{align}
	\end{subequations}
where $t_1=0$, and $\tilde{\psi}(x,t)$ and $\tilde{\chi}(x,t)$ represent the two spinor components of the exact moving soliton given by the ansatz \eqref{eq:psia}-\eqref{eq:chia}. For the ABS equation, the initial conditions $u_k^1$ and $v_k^1$ are simply discrete versions of the functions \eqref{eq26a}-\eqref{eq26b}, \ie $u_k^1=u(x_k,t_1)$ and $v_k^1=v(x_k,t_1)$.

\begin{figure}[h]
	\includegraphics[width=\linewidth]{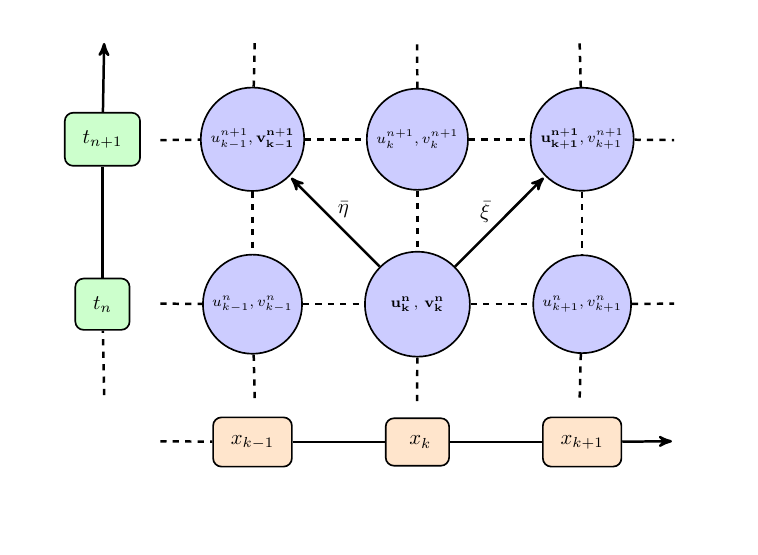}
	\caption{Sketch of the method of characteristics applied to solve the nonlinear Dirac equation. The numerical solutions $u_{k}^{n}$ and $v_{k}^{n}$ fulfill the Eqs.\ \eqref{eq:uf}-\eqref{eq:vf}. Starting from $u_{k}^{n}$ and $v_{k}^{n}$, the equations are integrated along the characteristic lines $\bar{\xi}$ 
	and $\bar{\eta}$, respectively. This procedure enables the solution  $u_{k+1}^{n+1}$ and $v_{k-1}^{n+1}$ to be obtained. }
	\label{fig:dia}
\end{figure}
The position of the center of the soliton is computed by the first moment of the charge 
normalized by the charge itself, \ie 
\bq
	\label{eq:posnum}
	q(t)=\dfrac{1}{Q}\int_{-\infty}^{+\infty}dx\, \rho(x,t) \,x.  
\eq
By deriving this expression with respect to time, by taking into account the 
continuity equation for the charge \eqref{eqc:charge}, and subsequent  to the integration by parts, the following definition  is obtained for the velocity
\bq
\label{eq:velnum}
\dot{q}(t)=\dfrac{1}{Q}\int_{-\infty}^{+\infty}dx\, \,j(x,t),  
\eq
where the current density is $j(x,t)=\chi\,\psi^\star+\chi^\star \,\psi$ for the GN model and $j(x,t)=|u|^2-|v|^2$ for the ABS model.

Here in after, the GN and ABS equations under linear and harmonic potentials are numerically solved by using Lakoba's method.

\section{Numerical simulations} \label{sec5}

\subsection{GN solitary wave under linear potential} 

  We consider the ramp potential $V(x)=-V_1 x$ with $V_1>0$. The simulation parameters are $\Delta x=\Delta t=h$ and $x\in [-L,L]$,  where $L$ has been properly chosen according to the soliton mobility and to the soliton width for each simulation. 
  The initial condition is represented by a GN soliton given by \eqref{eq:ic}  with velocity $\dot{q}(0)$, position $q(0)$, momentum $p(0)$, and frequency $\omega$. Let us start by analyzing the convergence, when $h$ approaches zero, of the numerical method for the perturbed GN equation \eqref{eq:sys} with parameters $\kappa=m=1$, $\omega=0.9$, and $V_1=10^{-2}$. The last value is the largest value for the parameter $V_1$, 
  which has been taken from Ref. \cite{mertens:2012}. Values larger than $10^{-2}$ cannot be considered as perturbative terms when the solitary wave leaves the origin. 
 
  In order to quantitatively investigate the convergence of the numerical method, let us define 
 the time evolution of the relative error corresponding to the magnitude ${\mathcal{A}}$ as 
 \begin{equation} \label{eq:error}
 	\epsilon[{\mathcal{A}}(t)]=\dfrac{\left|\mathcal{A}(t,h)-\mathcal{A}_{e}(t)\right|}{\mathcal{A}_{e}(t)} \text{,}
 \end{equation} 
 where $\mathcal{A}(t,h)$ is the computed quantity for a given $h$ at a time $t$, and 
 $\mathcal{A}_{e}(t)$ is the exact magnitude. 
 Therefore, the value of this residual error can be fixed, and simulations can be carried out in such a way that the difference between the calculated quantity and its exact magnitude never exceeds this value. When exact magnitude of $\mathcal{A}$ is unknown, the error \eqref{eq:error} is   computed through an approximated value $\tilde{\mathcal{A}}(t)$ calculated by means of the collective coordinate theory, instead of the exact value $\mathcal{A}_e(t)$.  In addition, the Simpson integration rule is used in order to compute the spatial integrals related with the charge, energy,  
 momentum, position, and velocity.
  
    In the \autoref{fig:lin_w_9e-1_V1_1e-2}(a)-\ref{fig:lin_w_9e-1_V1_1e-2}(b), the soliton position and momentum are represented, respectively,  for the following numerical steps: 
     $h=10^{-1}$, $10^{-2}$, $10^{-3}$, and $10^{-4}$.  
    The results are very similar for the different steps. The solitary wave is accelerated, while its momentum shows linear behavior in time.  
     The exact value of the momentum is $P_{e}(t)=V_1\,Q\,t$, whereas the approximated value of $q(t)$ is given by Eq. \eqref{eq:solq}. 
    	After  $t=70$, a slight difference arises among the momentum computed for $h=10^{-1}$ and those for the smaller values of $h$. 
    	As $h$ 
    approaches zero, it is observed that the 
    momentum tends towards its exact value. The same behavior is observed when the relative errors of the charge and the energy are plotted (see \autoref{fig:lin_w_9e-1_V1_1e-2}(c)-\ref{fig:lin_w_9e-1_V1_1e-2}(d)). Taking into account these results, henceforth we set $h=10^{-3}$. 
 
  \begin{figure}[h!]
  	\includegraphics[width=0.99\linewidth]{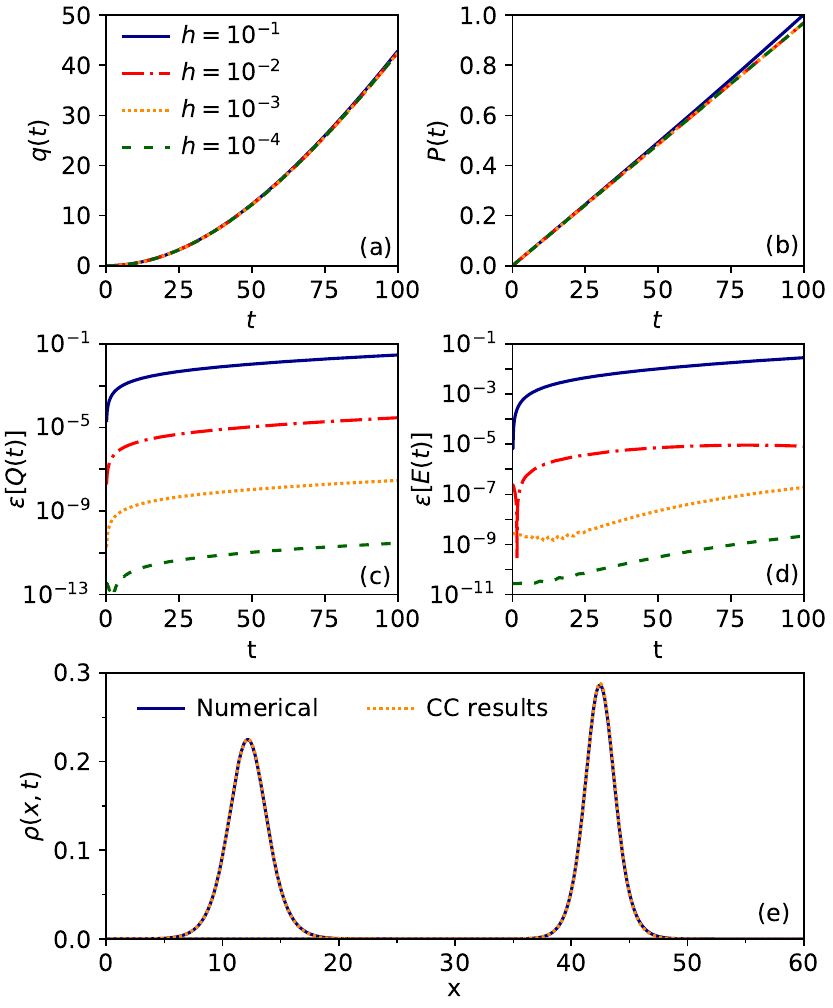}
  	\caption{Simulations of GN model ($m=1$ and $\kappa=1$) \eqref{eq:sys} with ramp potential, $V(x)=-V_1 x$, with $V_1=10^{-2}$. 
  	Time evolution of (a) position, (b) momentum,   
  	and the relative errors of (c) charge and (d) energy of the soliton from numerical simulations with step-size fixed at $h=10^{-1}$ (solid dark-blue line), $h=10^{-2}$ (dash-dotted red line), $h=10^{-3}$ (dotted orange line), and $h=10^{-4}$ (dashed dark-green line). (e) Comparison of soliton charge density, $\rho(x,t)=|\psi(x,t)|^2+|\chi(x,t)|^2$, at fixed times $t=50$ and $t=100$, obtained from numerical simulations with step-size fixed at $h=10^{-3}$ (solid dark-blue line) with those of the collective coordinate theory (dotted orange line). 
  		 The simulated length is $L=100$ and the initial condition uses the ansatz \eqref{eq:psia}-\eqref{eq:chia} with $q(0)=0$, $\dot{q}(0)=0$, $p(0)=0$, and $\omega=0.9$.
  }
\label{fig:lin_w_9e-1_V1_1e-2}  	
  \end{figure}  

 In the 
 \autoref{fig:lin_w_9e-1_V1_1e-2}(e), the density of the charge is shown as a function of  the spatial coordinate $x$ for fixed times. The soliton profiles computed from the simulations (solid dark-blue line) perfectly agree with the profiles obtained from the collective coordinate theory (dotted orange line).
  Lorentz contraction  of the soliton width due to the acceleration  is also shown.  
Simultaneously, the height of the soliton is increased owing to the conservation of the charge and no tails are developed in the soliton profiles. Thus, no soliton instability is observed. 
A longer integration time requires an extended domain length, which implies larger values of the external potential.

 To proceed with the analysis of the numerical stability of solitary waves under external potential, 
 a lower value of frequency,  $\omega=0.3$, is chosen. For this frequency, the solitary waves exhibits two humps, and the shape of the wave is deformed for $V_1=10^{-2}$ and is eventually destroyed (see Fig. 9  of Ref.\ \cite{mertens:2012}). Decreasing 
  the value of $V_1$ 
 by two orders of magnitude,
 only delayed the appearance of instabilities (see Fig. 10 of Ref.\ \cite{mertens:2012}).
 In contrast to these results, although a deformation of the two humps is observed for $V_1=10^{-2}$, the solitary waves here are not destroyed and both  relative errors of the energy and of the charge remain less than $10^{-6}$  (\autoref{fig:lin_w_3e-1_V1_1e-2}(a)-\ref{fig:lin_w_3e-1_V1_1e-2}(c)). Moreover, a detailed study of the evolution of the density of the charge shows that the amplitude of the two humps not only grows, but also oscillates with a frequency approximately equal to $2 \omega=0.6$. This fact use to be a precursor to the wave becoming unstable. Here the solitary wave is stable up to the final time of integration $t=100$.   
  Although the position of the center of mass, as well as the soliton 
 momentum, agree with the predicted values in Sec. \ref{sec3} (see  \autoref{fig:lin_w_3e-1_V1_1e-2}(d)-\ref{fig:lin_w_3e-1_V1_1e-2}(e)), the oscillations of the two humps of the charge density  cannot be described by our collective coordinate approach. This is due to the fact that the ansatz \eqref{eq:psia}-\eqref{eq:chia}, by construction, makes the charge density symmetric with respect to its center of mass.  
 
 \begin{figure}[h!]
 	\includegraphics[width=0.99\linewidth]{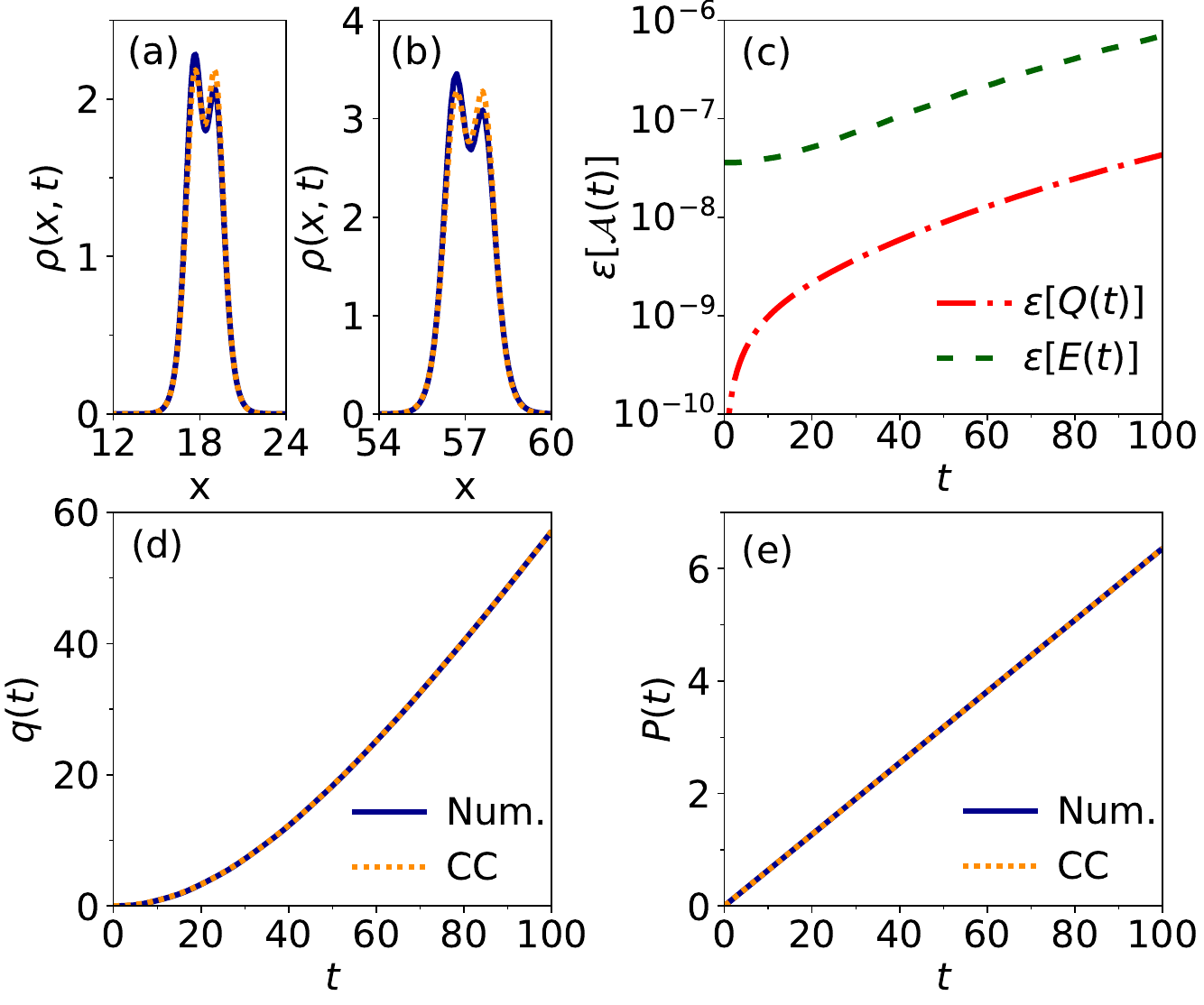}  
 	\caption{Simulations of GN model~($m=1$ and $\kappa=1$) \eqref{eq:sys} with ramp potential, $V(x)=-V_1 x$, with $V_1=10^{-2}$. 
 		Comparison of soliton charge density, $\rho(x,t)=|\psi(x,t)|^2+|\chi(x,t)|^2$, at fixed times (a)~$t=50$ and (b)~$t=100$, obtained from numerical simulations (solid dark-blue line) with those of the collective coordinate theory (dotted orange  line). 
 		(c) Time evolution of the relative errors of the soliton charge (dash-dotted red line) and energy (dashed dark-green line). Comparison of (d) position and (e) momentum of the soliton obtained from numerical simulations (solid dark-blue line) and from the collective coordinate theory (dotted orange line). The numerical  step-size is fixed at $h=10^{-3}$, the simulated length is $L=100$,  and	the initial condition uses the ansatz \eqref{eq:psia}-\eqref{eq:chia} with 
 		$q(0)=0$, $\dot{q}(0)=0$, $p(0)=0$, and $\omega=0.3$.
 	}
 	\label{fig:lin_w_3e-1_V1_1e-2}
 \end{figure}

 For the parameters considered in Fig. 10  in Ref.\ \cite{mertens:2012}, in which $V_1$ is decreased by two orders  of magnitude, the instabilities were reported  for  $t \approx 150$. Our simulations (\autoref{fig:lin_w_3e-1_V1_1e-4}) show stable waves even for larger value of the final time of integration. 
 In this case, the two-hump wave propagates practically without deformation and the ansatz describes perfectly not only the evolution of the collective variable, but also the soliton profile.

 \begin{figure}[h!]
 	\includegraphics[width=0.99\linewidth]{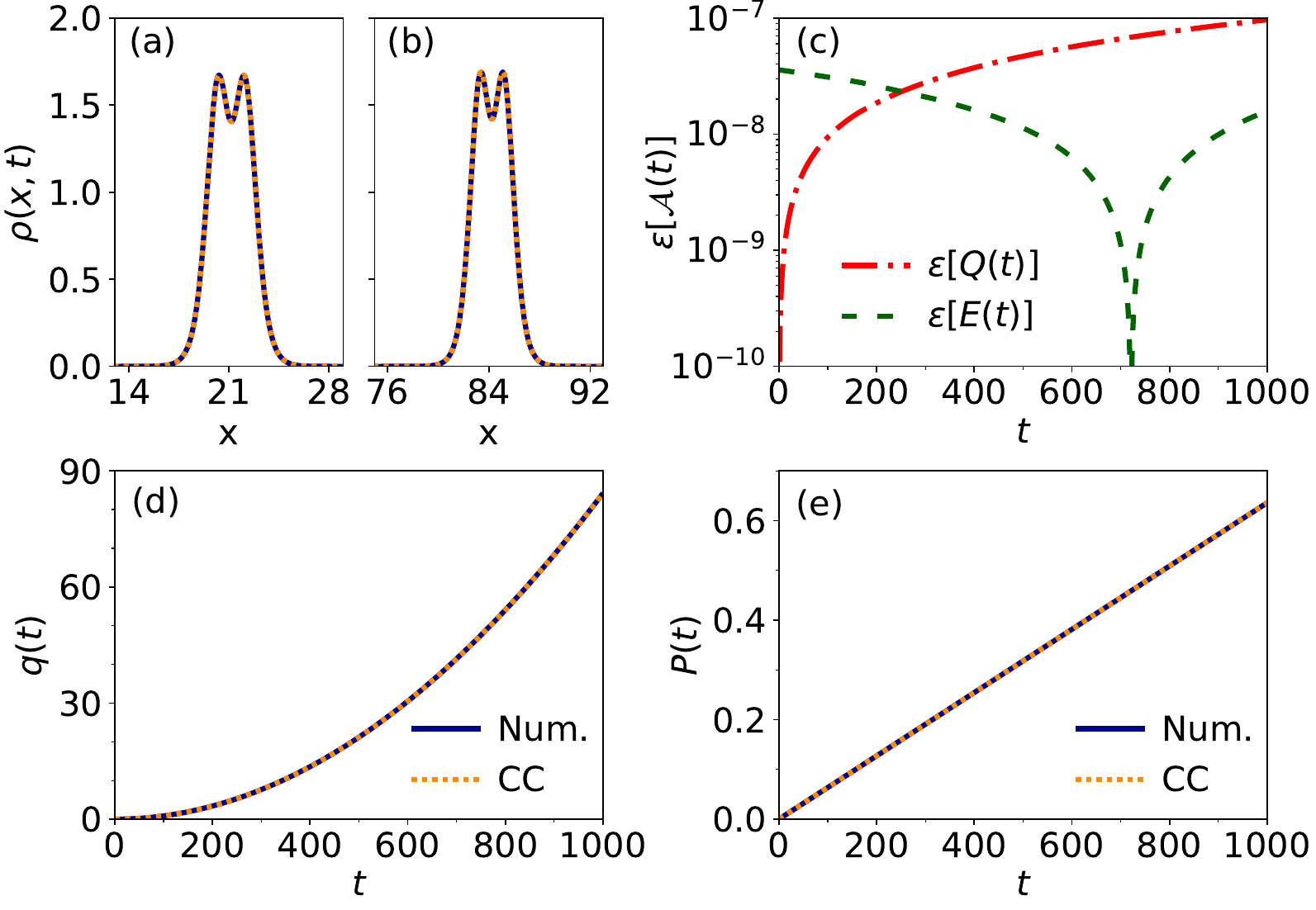}
 	\caption{Simulations of GN model~($m=1$ and $\kappa=1$) \eqref{eq:sys} with ramp potential, $V(x)=-V_1 x$, with $V_1=10^{-4}$. 
 		Comparison of soliton charge density, $\rho(x,t)=|\psi(x,t)|^2+|\chi(x,t)|^2$, at fixed times (a) $t=500$ and (b) $t=1000$, obtained from numerical simulations (solid dark-blue line) with those of the collective coordinate theory (dotted orange line). 
 		(c) Time evolution of the relative errors of the soliton charge (dash-dotted red line) and energy (dashed dark-green line). Comparison of (d) position and (e) momentum of the soliton obtained from numerical simulations (solid dark-blue line) and from the collective coordinate theory (dotted orange line). The numerical step-size is fixed at $h=10^{-3}$, the simulated length is $L=150$, and the initial condition  
 		 uses the ansatz \eqref{eq:psia}-\eqref{eq:chia} with $q(0)=0$, $\dot{q}(0)=0$, $p(0)=0$, and $\omega=0.3$. 
 	}
 	\label{fig:lin_w_3e-1_V1_1e-4}
 \end{figure}

   We have also explored the soliton's stability for lower values of $\omega$, when the two humps are more remarkable.  A 
  	lower value of $V_1$ makes the soliton slower and therefore the time of integration can be increased 
  up to $t=3000$ without the appearance of instability, see \autoref{fig:lin_w_1e-1_V1_1e-5}.

  \begin{figure}[h!] 
  	\includegraphics[width=0.99\linewidth]{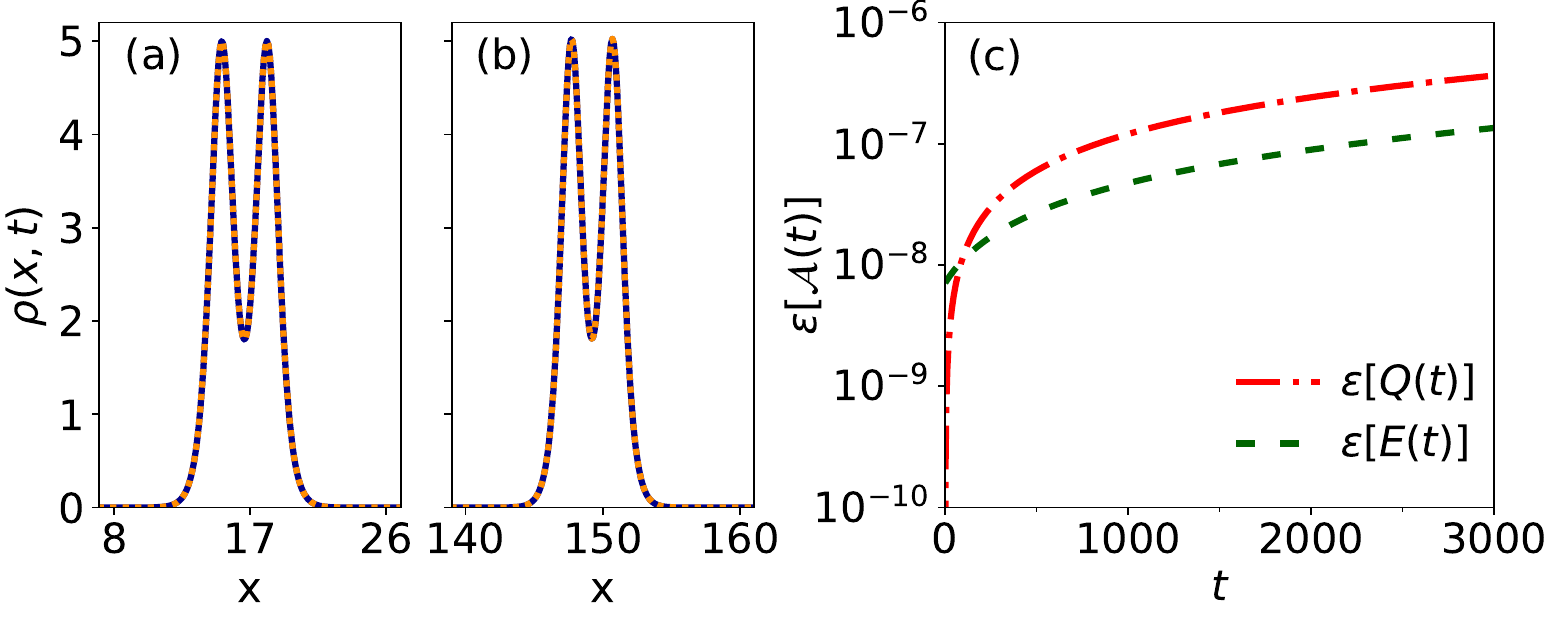}
  	\caption{Simulations of GN model ($m=1$ and $\kappa=1$) \eqref{eq:sys} with ramp potential, $V(x)=-V_1 x$, with $V_1=10^{-5}$. Comparison of soliton charge density, $\rho(x,t)=|\psi(x,t)|^2+|\chi(x,t)|^2$, at fixed times (a) $t=1000$ and (b) $t=3000$, obtained from numerical simulations (solid dark-blue line) with those of the collective coordinate theory (dotted orange  line). 
  		(c) Time evolution of the relative errors of the soliton charge (dash-dotted red line) and energy (dashed dark-green line).  
  		The numerical step-size is fixed at $h=10^{-3}$, the simulated length is $L=200$, and the initial condition  
  		 uses  the ansatz \eqref{eq:psia}-\eqref{eq:chia} with $q(0)=0$, $\dot{q}(0)=0$, $p(0)=0$, and $\omega=0.1$.
  	}
  	\label{fig:lin_w_1e-1_V1_1e-5}
  \end{figure}

   Since the soliton under the linear potential accelerates, a larger integration time requires a larger spatial domain. On the contrary, the dynamics of the confined soliton can be studied over a longer integration time, once the length of the system is fixed. It is precisely the behavior that the soliton exhibits in the harmonic potential.

\subsection{GN solitary wave under harmonic potential}

  When a harmonic potential $V(x)=(V_2/2) x^2$ with $V_2>0$ is considered, if the center of the 
  soliton is initially  at the minimum of the potential $q(0)=0$, then is necessary a non-zero initial velocity  so that the solitary wave moves. This is the reason why  $\dot{q}(0)=0.1$ (non-relativistic case) and $\dot{q}(0)=0.9$ (relativistic case) are chosen in our simulations. For the former case, the center of mass harmonically oscillates around its initial position, whereas for the latter cases, $q(t)$ oscillates an-harmonically. 
  
  According to the simulations of Ref. \cite{mertens:2012},  one-hump ($\omega=0.9$)  and two-hump  ($\omega=0.3$) solitary wave were deformed after a certain time. However, our  simulations 
  show that these waves are stable within the range of studied parameters. Notice that, in our simulations,  
  a longer integration time  than in Figures 12 and 13  of Ref. \cite{mertens:2012} is taken, and no insights  of instabilities are detected (see 
  Figs.~\ref{fig:harm_w_9e-1_V2_1e-4_v0_0.9} and \ref{fig:harm_w_3e-1_V2_1e-4_v0_0.1}).  

  The simulations of the relativistic soliton of \autoref{fig:harm_w_9e-1_V2_1e-4_v0_0.9} is too time consuming due to the needed long simulated length. In contrast, simulations of non-relativistic soliton of  \autoref{fig:harm_w_3e-1_V2_1e-4_v0_0.1} have been performed up to much longer times without excessive computational time, 
 	 and no instabilities are observed. For the final time  of simulation $t=10000$ (not shown in \autoref{fig:harm_w_3e-1_V2_1e-4_v0_0.1}), the maximal relative error of the charge and energy  are less than $10^{-6}$.

 \begin{figure}[h!]
	\includegraphics[width=0.99\linewidth]{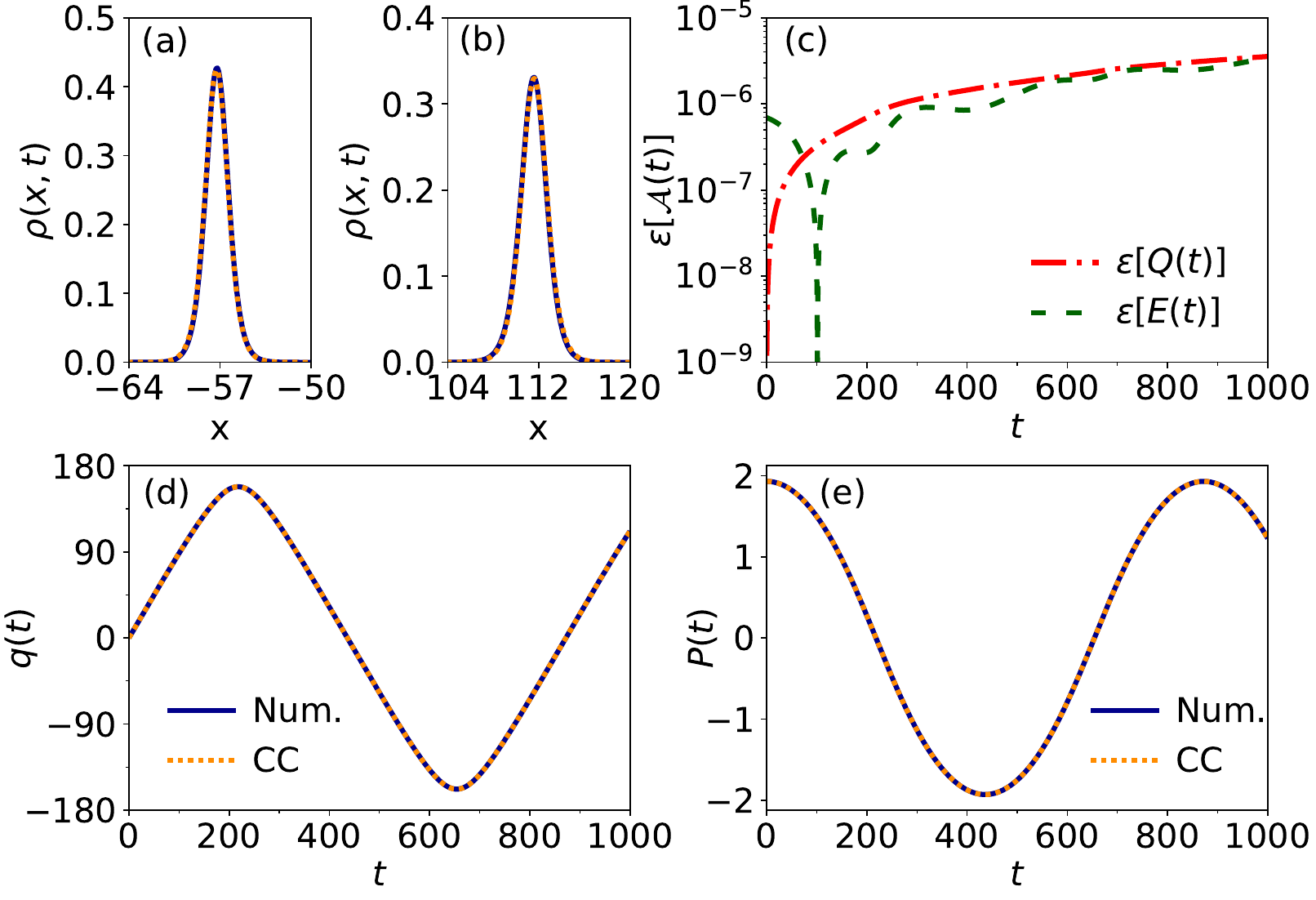}
	\caption{Simulations of GN model ($m=1$ and $\kappa=1$) \eqref{eq:sys} with harmonic potential, $V(x)=(V_2/2) x^2$, with $V_2=10^{-4}$.
		 Comparison of soliton charge density, $\rho(x,t)=|\psi(x,t)|^2+|\chi(x,t)|^2$, at fixed times (a) $t=500$ and (b) $t=1000$, obtained from numerical simulations (solid dark-blue line) with those of the collective coordinate theory (dotted orange line). 
		(c) Time evolution of the relative errors of the soliton charge (dash-dotted red line) and energy (dashed dark-green line).
		 Comparison of (d) position  and (e) momentum of the soliton obtained from numerical simulations (solid dark-blue line) and from the collective coordinate theory (dotted orange line). The numerical step-size is fixed at $h=10^{-3}$, the simulated length is $L=300$, and the initial condition 
		 uses  the ansatz \eqref{eq:psia}-\eqref{eq:chia} 
		with $q(0)=0$, $\dot{q}(0)=0.9$,  $p(0)=1.86$, 
		and $\omega=0.9$. 
	}
	\label{fig:harm_w_9e-1_V2_1e-4_v0_0.9}
\end{figure}

\begin{figure}[h!]
	\includegraphics[width=0.99\linewidth]{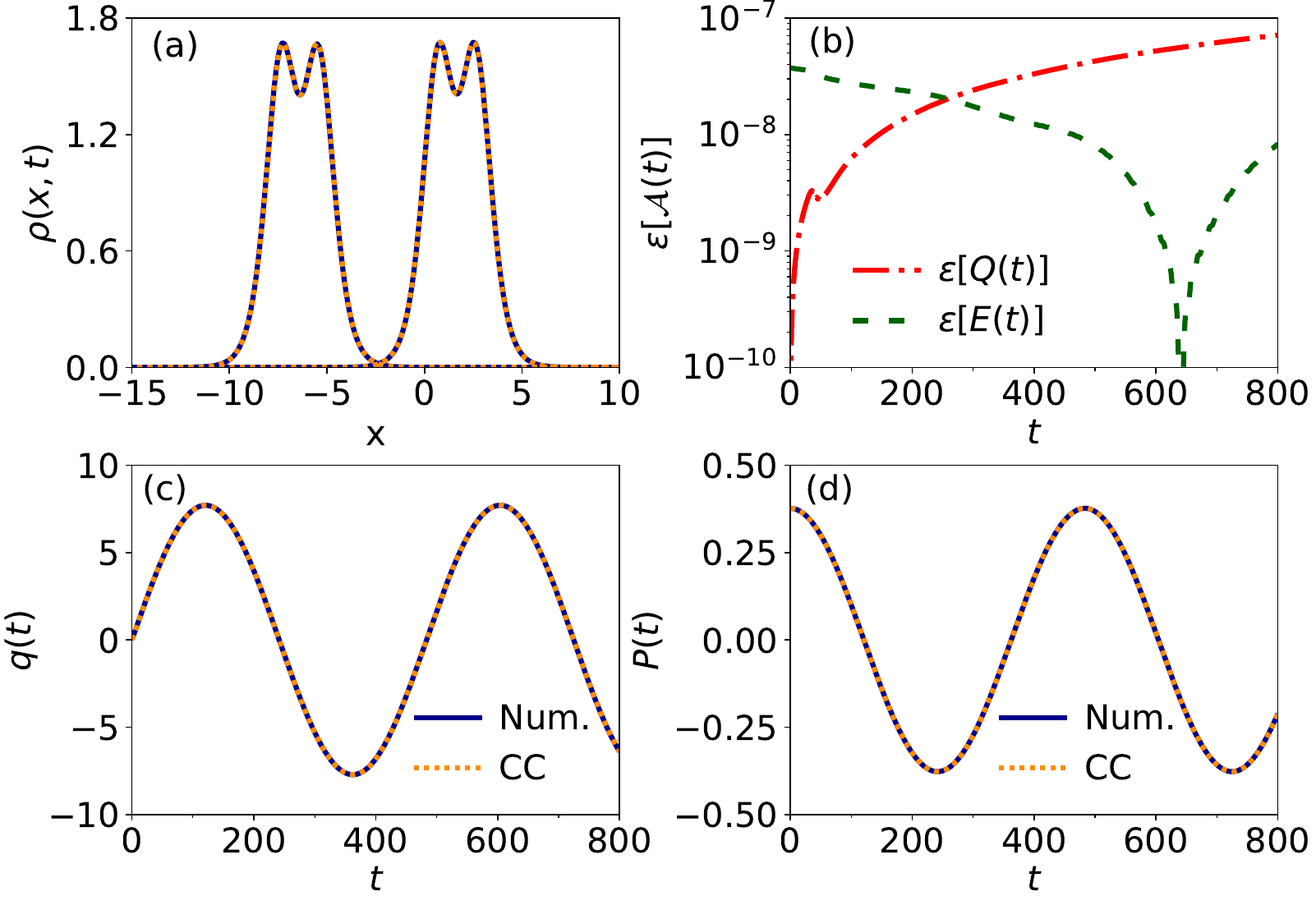} \\
	\caption{Simulations of GN model ($m=1$ and $\kappa=1$) \ \eqref{eq:sys} with harmonic potential, $V(x)=(V_2/2) x^2$, with $V_2=10^{-4}$.  
		Comparison of soliton charge density, $\rho(x,t)=|\psi(x,t)|^2+|\chi(x,t)|^2$, at fixed times  (a) $t=500$ and $t=800$, obtained from numerical simulations (solid dark-blue line) with those of the collective coordinate theory (dotted orange  line).
		(b) Time evolution of the relative errors of the soliton charge (dash-dotted red line) and energy (dashed dark-green line). Comparison of (c) position and (d) momentum of the soliton obtained from numerical simulations (solid dark-blue line) and from the collective coordinate theory (dotted orange line). The numerical step is fixed at $h=10^{-3}$, $L=30$, the initial condition  uses  the ansatz \eqref{eq:psia}-\eqref{eq:chia} with $q(0)=0$,  $\dot{q}(0)=0.1$, $p(0)=0.03$, and $\omega=0.3$. 
	}
	\label{fig:harm_w_3e-1_V2_1e-4_v0_0.1}
\end{figure}

   For a soliton with frequency $\omega=0.1$ in the same harmonic potential, $V_2=10^{-4}$, the initial shape of the soliton is slightly distorted, see \autoref{fig:GN_w_0.1_ABC}(a), 
   however, the errors of the charge and energy grow up to almost $0.1$ (see the dash-dotted red line and the dashed green line in \autoref{fig:GN_w_0.1_ABC}(b)) which is an indication of instability. These instabilities, as in the case of the GN equation without potential can be removed by using the so-called absorbing boundary conditions (see the solid blue line and the dotted orange line in \autoref{fig:GN_w_0.1_ABC}~(b)),  
  	that is, the numerical solution, each $h_{abc}$ interval of time, is multiplied by the following function 
  	\begin{equation}
  		\label{eq:abc} 
  		\rho_a(x)=\exp\left[-\left(\frac{|x|-L_1}{W}\right)^2\right], 
  	\end{equation}  
  if $|x| \in [L_1,L]$, where $W=(L-L_1)/B$, and $L_1<L$ and $B$ are parameters. 
  	For more details on how to select these parameters, see Ref. \cite{lakoba:2018}. One disadvantage of using the absorbing boundary conditions is that it affects the soliton position. In 
  \autoref{fig:GN_w_0.1_ABC}(a) it may be appreciated the differences among the soliton profiles obtained from the collective coordinate approach, and the two profiles obtained from simulations with and without absorbing boundary conditions (ABC).

\begin{figure}[h!]   
	\includegraphics[width=0.99\linewidth]{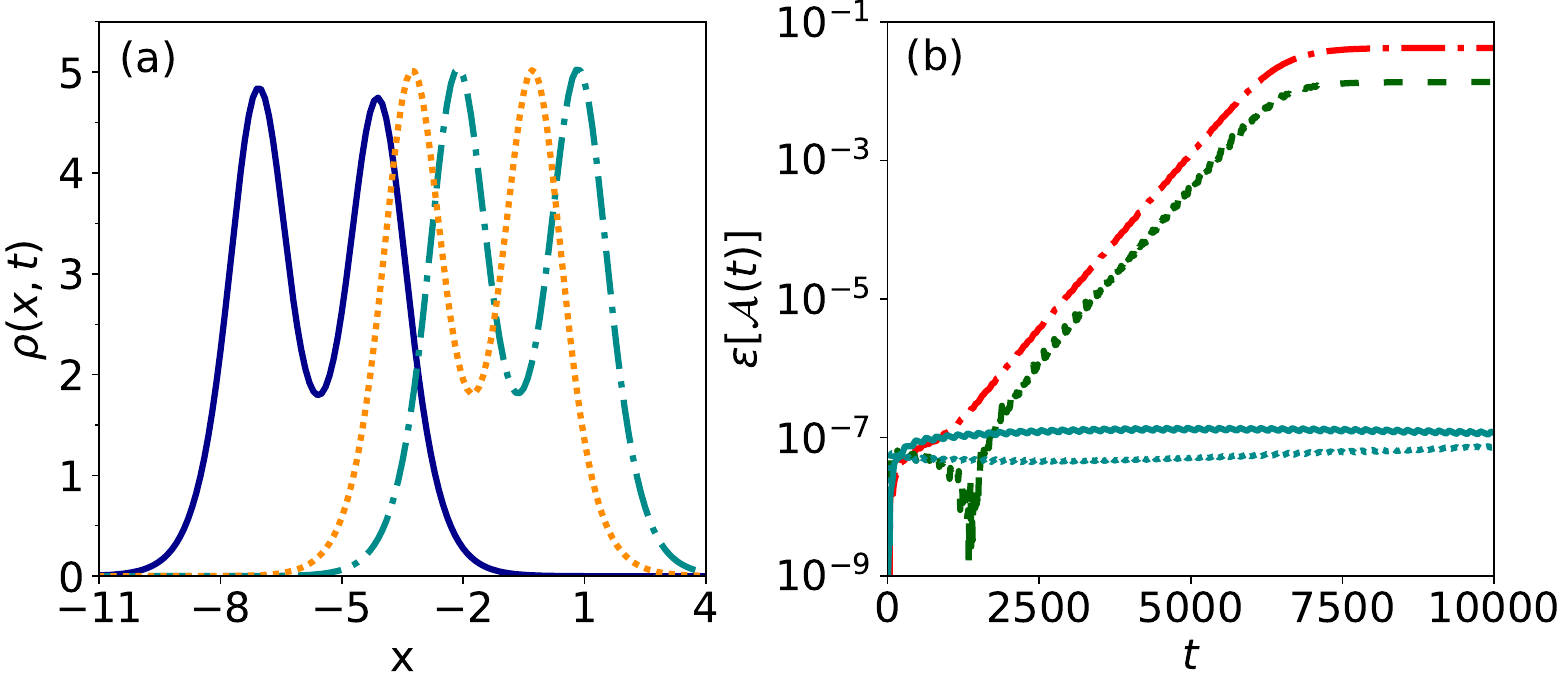}
	\caption{Simulations of GN model~($m=1$ and $\kappa=1$) \eqref{eq:sys} with harmonic potential, $V(x)=(V_2/2) x^2$, with $V_2=10^{-4}$ and with and without absorbing boundary conditions.  
		(a) Comparison of soliton charge density, $\rho(x,t)=|\psi(x,t)|^2+|\chi(x,t)|^2$, at fixed time $t=10000$ obtained from the simulations without  ABC (solid dark-blue line) and 
		with ABC (dash-dotted dark-cyan line), with that of the collective coordinate theory (dotted orange  line). (b) Relative errors of the soliton charge without ABC (dash-dotted red line), with ABC (solid dark-cyan line) and energy without ABC (dashed dark-green line) and with ABC (dotted dark-cyan line).
		The numerical step is fixed at $h=10^{-3}$, the simulated length is $L=30$, the initial condition  uses the ansatz \eqref{eq:psia}-\eqref{eq:chia} with $q(0)=0$, $\dot{q}(0)=0.1$,  
		$p(0)=0.01$,  
		and $\omega=0.1$, 
		and the parameters of the absorbing boundary conditions \cite{lakoba:2018} are $L_1=0.4 \cdot{L}$, $B=0.05$, and $h_{abc}=0.4$.}			
	\label{fig:GN_w_0.1_ABC}
\end{figure}

  	A similar behavior is observed for all frequencies lower than $0.1$, see \autoref{fig:GN_V2_1e-4_barrido_w_L_100}, where it is studied the time evolution of the relative errors of the charge and  energy  as a function of the frequency, $\omega \in [0.01,0.99]$. 
  	 In order for the errors to be of the order of $10^{-6}$ or less, 
   we search for the parameters $B$, $L_1$, and $h_{abc}$ of ABC which minimize the maximal errors of both charge and energy, for each low-frequency soliton.
 The optimal ABC parameters are collected in Tables \ref{tab2} and \ref{tab2a}. It is worth noting that the values of $L_1$ and $B$ that minimize the errors when $0.02\le \omega \le 0.1$ do not when $\omega=0.01$. In 
  	fact, the parameters in Table \ref{tab2a} show that a small variation of the values of $L_1$, $B$, and $h_{abc}$, may vary the errors some orders of magnitude. Furthermore, the errors are significantly reduced when $L_1=0.3614L$ is chosen instead of $L_1=0.4\,L$, while the errors hardly vary when $B$ and $h_{abc}$ change slightly.    	
  	Since the high-frequency solitons are broader, $L=100$ has been chosen 
  	 in \autoref{fig:GN_V2_1e-4_barrido_w_L_100}.

\begin{figure}[h!]
	\centering
	\begin{tabular}{c}
		\includegraphics[width=0.95\linewidth]
		{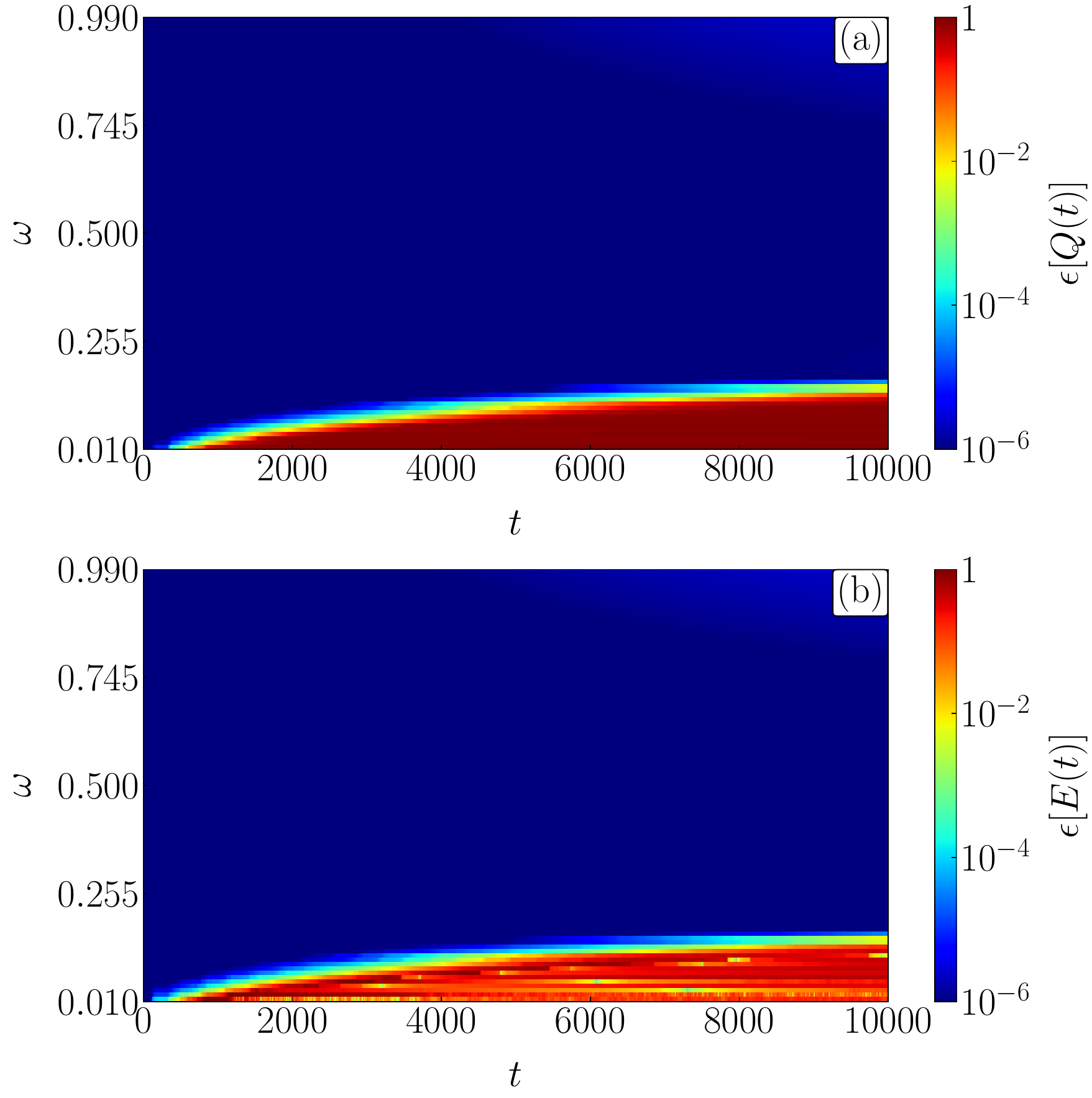}
	\end{tabular}
	\caption{Simulations of GN model~($m=1$ and $\kappa=1$) with harmonic potential, $V(x)=(V_2/2) x^2$, with $V_2=10^{-4}$.  
			Time evolution of the relative errors of the soliton (a) charge and (b) energy as a function of the frequency. 
		 The numerical step is fixed at $h=10^{-3}$, the simulated length is $L=100$, and the initial condition  uses  the ansatz \eqref{eq:psia}-\eqref{eq:chia} with $q(0)=0$, $\dot{q}(0)=0.1$,   
		and $p(0)$ satisfying Eq.~\eqref{eq:p}.}
	\label{fig:GN_V2_1e-4_barrido_w_L_100} 	
\end{figure}
 
 \begin{table}
 	\begin{tabular}{|c|c|c|c|}
 		\hline
 		$\omega$ & $h_{abc}$  & $max(\epsilon[Q])$     &                $ max(\epsilon[E])$ \\
		\hline 
 		$0.02$   &  $2.9 \times 10^{-2}$ & $4.83\times 10^{-6}$ &$1.45\times 10^{-6}$ \\
 		\hline  
 		$0.03$   &  $ 2.0\times 10^{-2}$ & $1.76\times 10^{-6}$ &$7.25\times 10^{-7}$ \\	
 		\hline
 		$0.04$   &  $2.6 \times 10^{-2}$ & $1.79\times 10^{-6}$ &$4.50\times 10^{-7}$ \\
 		\hline
 		$0.05$       & $9.0 \times 10^{-2}$ & $7.37\times 10^{-7}$ & $5.25 \times 10^{-7}$\\
 		\hline
 		$0.06$  & $9.0\times 10^{-2}$ & $1.08\times 10^{-6}$ & $5.69\times 10^{-7}$
 		\\
 		\hline
 		$0.07$  & $8.5\times 10^{-2}$ & $1.28\times 10^{-6}$& $1.87\times 10^{-7}$
 		\\
 		\hline
 		
 		$0.08$ & $1.2\times 10^{-1}$ & $1.30\times 10^{-6}$ & $2.45\times 10^{-7}$
 		\\
 		\hline
 		
 		$0.09$ & $2.0\times 10^{-1}$ & $6.91\times 10^{-7}$ &$9.79\times 10^{-8}$
 		\\
 		\hline
 		$0.10$ & $4.0\times 10^{-1}$ & $1.38\times 10^{-7}$ &$7.55\times 10^{-8}$ \\
 		\hline
 	\end{tabular}
 	\caption{Simulations  of GN Eqs.\ \eqref{eq:sys} with a harmonic potential, $V(x)=(V_2/2) x^2$. 
 	 For low frequencies, are shown the 
 			interval of times $h_{abc}$, which minimize the maximal values of relative errors of the charge and energy, when $L_1=0.4\,L$ and $B=0.05$. 
	 Parameters of simulations and initial conditions are $V_2=10^{-4}$, $h=10^{-3}$,  $L=30$, $q(0)=0$, $\dot{q}(0)=0.1$, and $p(0)$  satisfying \eqref{eq:p}.}
 	 \label{tab2}
 \end{table}   
 
 \begin{table}
	\begin{tabular}{|c|c|c|c|c|}
		\hline
		$L_1$ &	$B$ & $h_{abc}$  & $max(\epsilon[Q])$     &                $ max(\epsilon[E])$ \\
		\hline
$0.4\,L$	&	$0.05$   &  $1.9\times10^{-2} $ & $3.7\times 10^{-3}$ &$2.9\times 10^{-4}$ \\
		\hline 
$0.3614\,L$	&	$0.05$   &  $1.9\times10^{-2}$ & $2.8\times 10^{-5}$ &$7.3\times 10^{-6}$ \\
		\hline  
$0.3614\,L$	&	$0.0531$   &  $1.9\times10^{-2} $ & $2.8\times 10^{-5}$ &$7.1\times 10^{-6}$ \\
		\hline  
$0.3614\,L$	&	$0.0531$   &  $2.0\times10^{-2} $ & $2.7\times 10^{-5}$ &$7.2\times 10^{-6}$ \\	
		\hline
	\end{tabular}
	\caption{ Simulations  of GN Eqs.\ \eqref{eq:sys} with a harmonic potential, $V(x)=(V_2/2) x^2$, and ABC. 
		Maximal values of relative errors of the charge and energy for $\omega=0.01$  
		for the given parameters $L_1$, $B$ and $h_{abc}$.
	Parameters and initial conditions are $V_2=10^{-4}$, $h=10^{-3}$,  $L=30$, $q(0)=0$, $\dot{q}(0)=0.1$, and $p(0)$ satisfying \eqref{eq:p}.}
	\label{tab2a}
\end{table}

\subsection{ABS solitary wave under linear and harmonic potentials}

  Finally, we investigate the stability of the solitary wave in the ABS model under linear and harmonic potentials. For the set of parameters considered  in Ref. \cite{mertens:2021}, we compute the numerical solutions of Eqs.\ \eqref{eq:sys} with $\kappa=0$ and $m=-1$ by using Lakoba's  methodology. For the linear potential,  the computed values of 
  $q(t)$, $\dot{q}(t)$, and $P(t)$ are compared with the exact analytical solution of the collective coordinate equations  
  \eqref{eq:solp}-\eqref{eq:solq}, see \autoref{fig:abslin_w_71e-2_V1_1e-2}(a)-(c). Although the equations of motion can be solved analytically, the solutions are only approximated due to the ansatz. Here, the function $P(t)$ also agrees with its exact solution. The relative errors of the charge and energy are always less than $10^{-6}$, see \autoref{fig:abslin_w_71e-2_V1_1e-2}(d), and can be decreased by taking lower values of $h$. 
  
In   \autoref{fig:abslin_w_71e-2_V1_1e-2}(e)  
 the solitary wave propagates to the right, and the two humps oscillate. These oscillations 
  are not described by the collective coordinate theory developed here, and can be the cause of the fluctuations of the numerical values of  $\dot{q}(t)$ and $q(t)$, which are not expected from the solution of its approximated values Eqs. \eqref{eq:solqdot}-\eqref{eq:solq}.

  In order to further investigate this issue, we additionally compute the velocity from the results of the soliton position by 
  approximating the first derivative by   
   means of $8^{\text{th}}$-order centered finite difference. 
    The value of $h$ is also reduced and the interval limits of the simulated spatial coordinate is increased. 
    Nevertheless, similar oscillations appear in all  simulations. Once we determine the convergence of the numerical solution for $h \le 10^{-3}$ and $L \ge 100$, two types of numerical  simulations are  performed: 
   in the first, the value of $V_1$ is reduced so that the hump oscillations hardly appear (see  \autoref{fig:abslin_w_71e-2_V1_1e-4}(a)), while in the second, $\omega$ is changed so that the two humps almost disappear (see \autoref{fig:abslin_w_71e-2_V1_1e-4}(c)). In both cases, the other parameters are fixed as in \autoref{fig:abslin_w_71e-2_V1_1e-2}.   
  The  \autoref{fig:abslin_w_71e-2_V1_1e-4}(b) shows the time evolution of the velocity for a lower value of $V_1=10^{-4}$, which agrees with the velocity obtained from the collective coordinate theory. In this case, the solitary wave still has two humps, see \autoref{fig:abslin_w_71e-2_V1_1e-4}(a), but its oscillations are practically imperceptible. By increasing the frequency up to $\omega=0.74$, the soliton profile practically has one hump and the computed soliton velocity also agrees with the expected result from the collective coordinate theory as is shown in  \autoref{fig:abslin_w_71e-2_V1_1e-4}(d).   Therefore, the 
  	amplitude of the oscillations of the velocity is related with the oscillations of the two humps.   
    
   \begin{figure}[h!]
  	\includegraphics[width=0.99\linewidth]{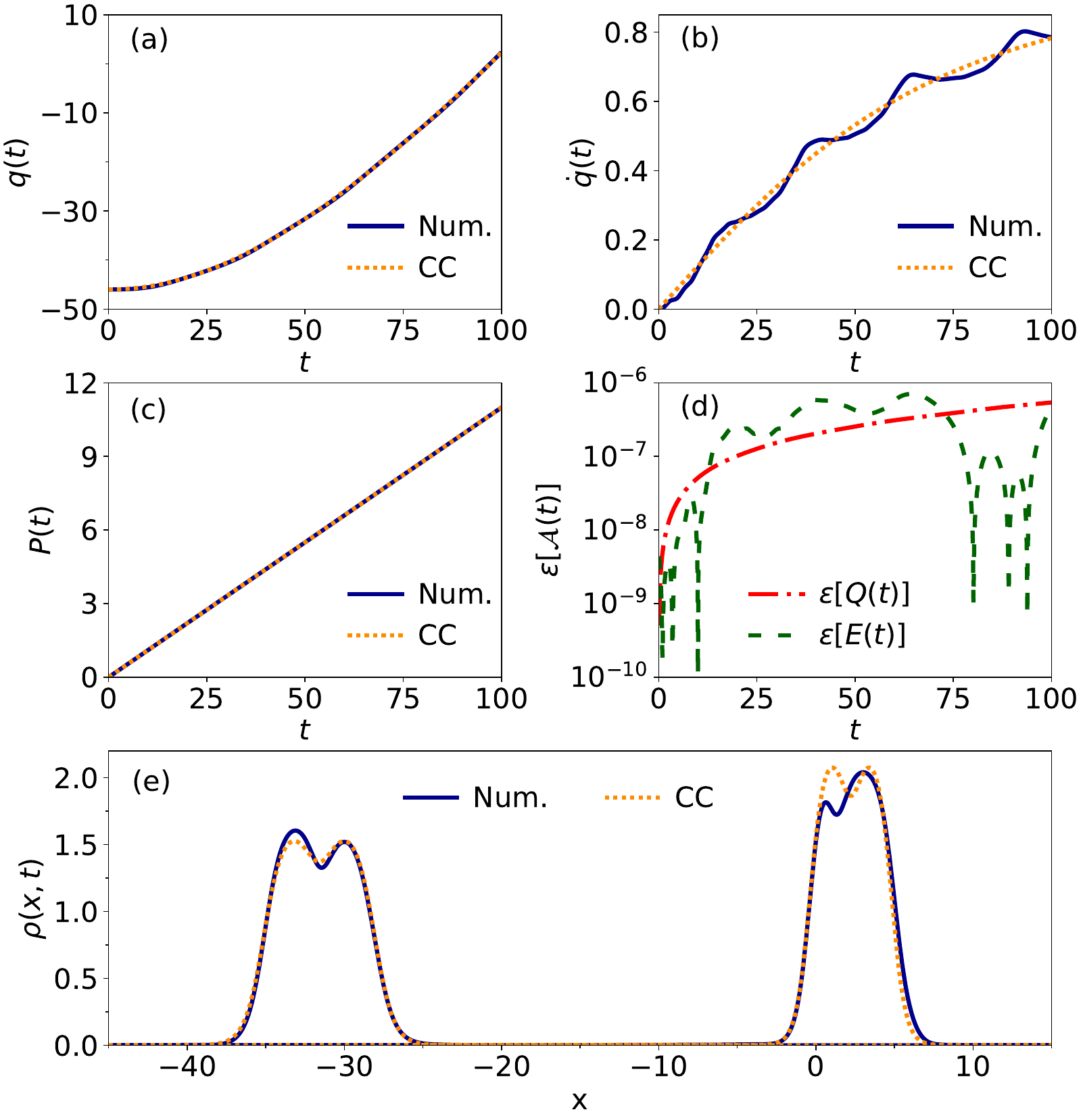}
  	\caption{Simulations of ABS model~($m=-1$ and $\kappa=0$) with ramp potential, $V(x)=-V_1 x$, with $V_1=10^{-2}$.  
  		Comparison of the time evolution of (a) soliton position, (b) soliton velocity, and (c) soliton momentum,  
  		  obtained from numerical simulations (solid dark-blue line) with the analytical results of the collective coordinate theory~\eqref{eq:sys} (dotted orange line).
  		(d) Time evolution of the relative errors of the soliton charge (dash-dotted red line) and energy (dashed dark-green line). 
  		(e) Comparison of soliton charge density, $\rho(x,t)=|u(x,t)|^2+|v(x,t)|^2$, at fixed times $t=50$ and $t=100$, obtained from numerical simulations (solid dark-blue line) with those of the collective coordinate theory (dotted orange line).   		
  		  The numerical step-size is fixed at $h=10^{-3}$, the simulated length is $L=150$, and the initial condition is the ansatz \eqref{eq26a}-\eqref{eq26b} with $q(0)=-46$, $\dot{q}(0)=0$, $p(0)=0$, and $\omega=0.71$. }
  	\label{fig:abslin_w_71e-2_V1_1e-2}
  \end{figure}
  
  \begin{figure}[h!]	
  	\includegraphics[width=0.99\linewidth]{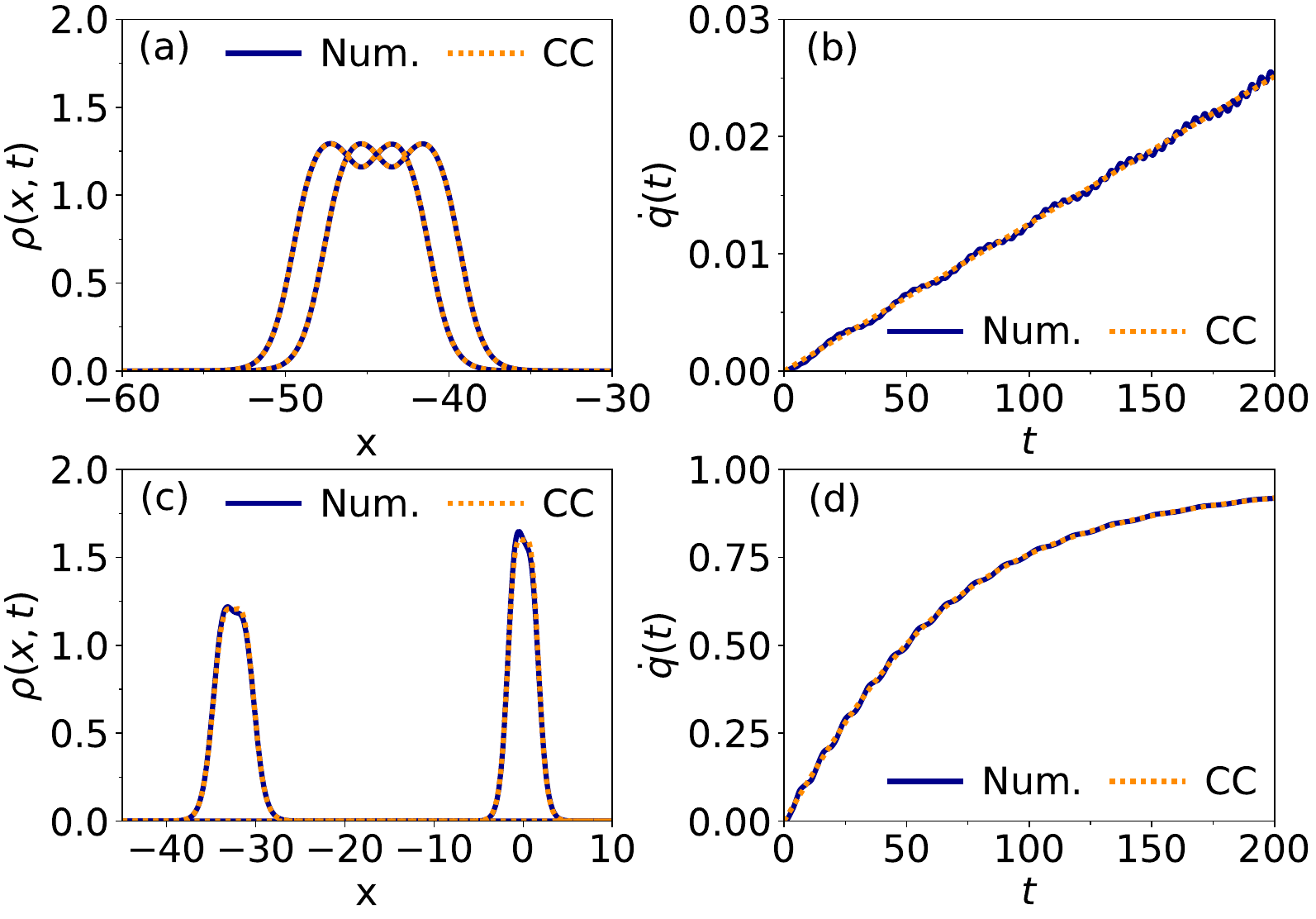}
	\caption{Simulations of ABS model~($m=-1$, and $\kappa=0$) with ramp potential, $V(x)=-V_1 x$. Comparison of (a) and (c) soliton charge density, $\rho(x,t)=|u(x,t)|^2+|v(x,t)|^2$, at fixed times $t=100$ and $t=200$, and of (b) and (d) soliton velocity, $\dot{q}(t)$, obtained from numerical simulations (solid dark-blue line) with those of analytical results from the collective coordinate theory (dotted orange line).  The numerical step-size is fixed at $h=10^{-3}$, the initial condition is the ansatz \eqref{eq26a}-\eqref{eq26b} with $q(0)=-46$, $\dot{q}(0)=0$,  and $p(0)=0$,    
		and the parameters of simulations are (a)-(b) $\omega=0.71$, $V_1=10^{-4}$, and  $x \in [-80,80]$, and (c)-(d) $\omega=0.74$, $V_1=10^{-2}$, and  $x \in [-150,150]$.}
	\label{fig:abslin_w_71e-2_V1_1e-4}
\end{figure}

 Let us emphasize that the stability of solitons of the ABS model 
	without perturbation was discussed in Ref. \cite{alexeeva:2019}, where the authors concluded that the solitons are stable, except in the 
	region of low frequency, in which $\omega \in {\cal{I}}_u=(1/\sqrt{2},0.729)$. 
	Outside this interval, i.e., $ \omega \in {\cal{I}}_s =(0.729, 1)$ the real part of the eigenvalues of the corresponding stability problem was found less than the critical value, $\lambda_{crit}=0.003$, and the soliton was defined as stable. 	
Note that the soliton can look stable for short time simulations as it is shown in \autoref{fig:abslin_w_71e-2_V1_1e-4}, even if the frequency $\omega \in {\cal{I}}_u$. 
In contrast, for large time simulations as those carried out when a harmonic potential is considered, if the real part of a given eigenvalue is,  for instance, $\lambda_r=0.1 \cdot \lambda_{crit}=3 \times 10^{-4}$, the numerical solution of spinors 
	will grow as $\exp(\lambda_r\,t)$ and the soliton will be unstable because the final time of integration is of order of $10^4$. Therefore, the upper frequency of the interval ${\cal{I}}_u$ 
	depends on  
	$\lambda_{crit}$ and can be slightly greater than $0.729$.  
	
\begin{figure}[h!]
  	\includegraphics[width=0.99\linewidth]{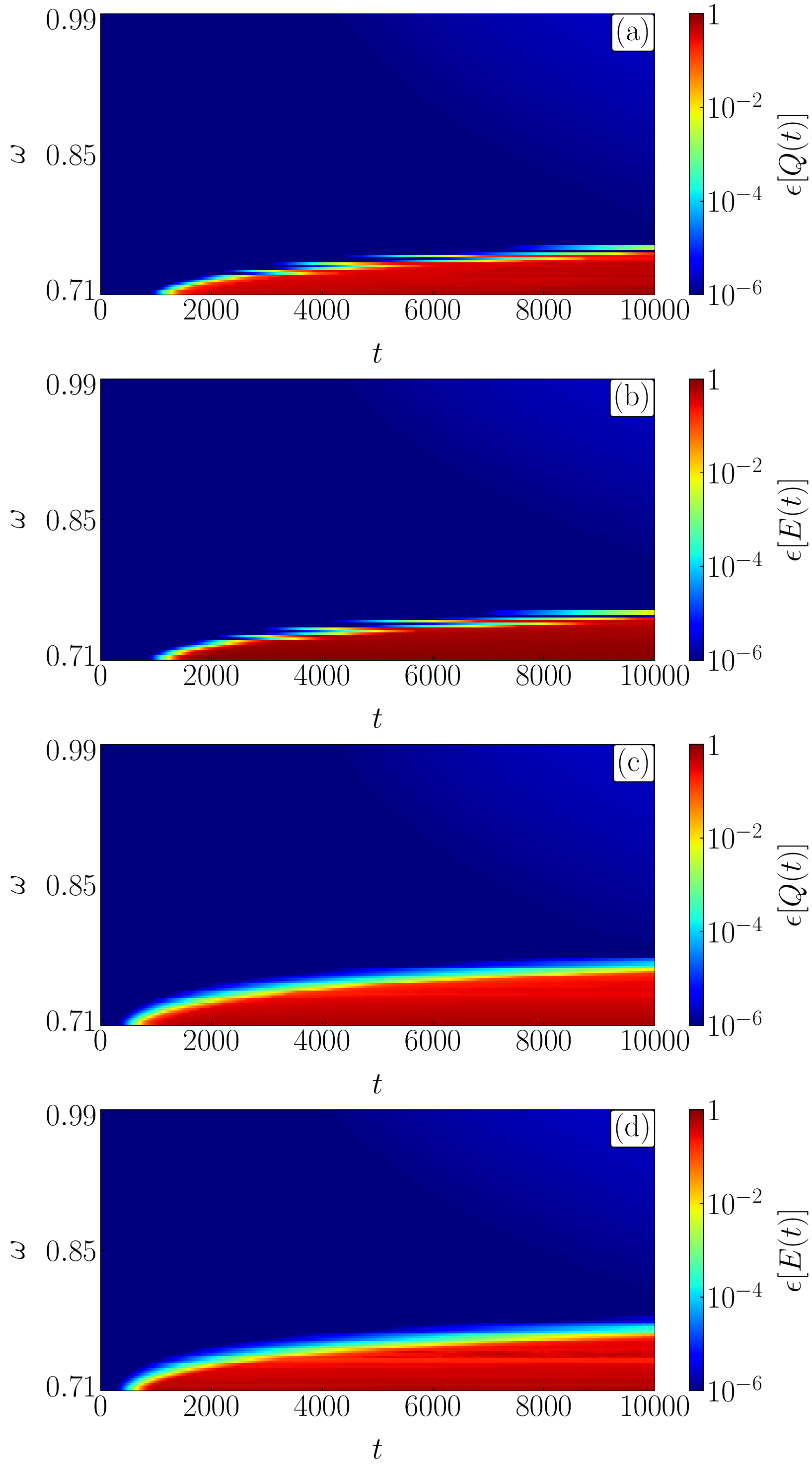}

	\caption{Simulations of ABS model~($m=-1$ and $\kappa=0$) (a)-(b) without and (c)-(d) with harmonic potential,  $V(x)=(V_2/2) x^2$, with $V_2=10^{-4}$. Time evolution of the relative errors of the soliton (a) and (c) charge, and (b) and (d) energy, as a function of the frequency. The numerical step is fixed at $h=10^{-3}$, the simulated length is $L=100$, and the initial condition is  the ansatz \eqref{eq26a}-\eqref{eq26b} with $q(0)=0$, $\dot{q}(0)=0.1$, and   
			$p(0)$ satisfying Eq.\ \eqref{eq:p}.}
	\label{fig:ABS_V2_1e-4_barrido_w_L_100} 	
\end{figure}

   	Our simulations of the ABS model without potential confirms 
   	this conjecture. 
 Indeed, we show that the charge and the energy errors are of order of $10^{-4}$ or greater than this value for  $\omega \in [0.71,0.76]$ 
  in  \autoref{fig:ABS_V2_1e-4_barrido_w_L_100}(a)-\ref{fig:ABS_V2_1e-4_barrido_w_L_100}(b). 
	The higher the frequency within this interval, the later the soliton destabilizes. 
	In this case, the one-hump ($\omega \in (0.75,0.76)$) or the two-humps ($\omega \in (1/\sqrt{2},0.75)$) of the stationary nonlinear wave, oscillate. The amplitude of these oscillations increases with time while radiation appears, until the wave is destroyed. Here, the two humps of the solitary wave oscillate with the same phase (see \autoref{fig:13}(a)). 
	In the presence of potential, the oscillations of the humps appear early. They are no longer symmetric with respect to the center of the mass since the wave moves in a spatial harmonic potential (see \autoref{fig:13}(b)).     

\begin{figure}[h!]
	\centering
	\includegraphics[width=0.95\linewidth]{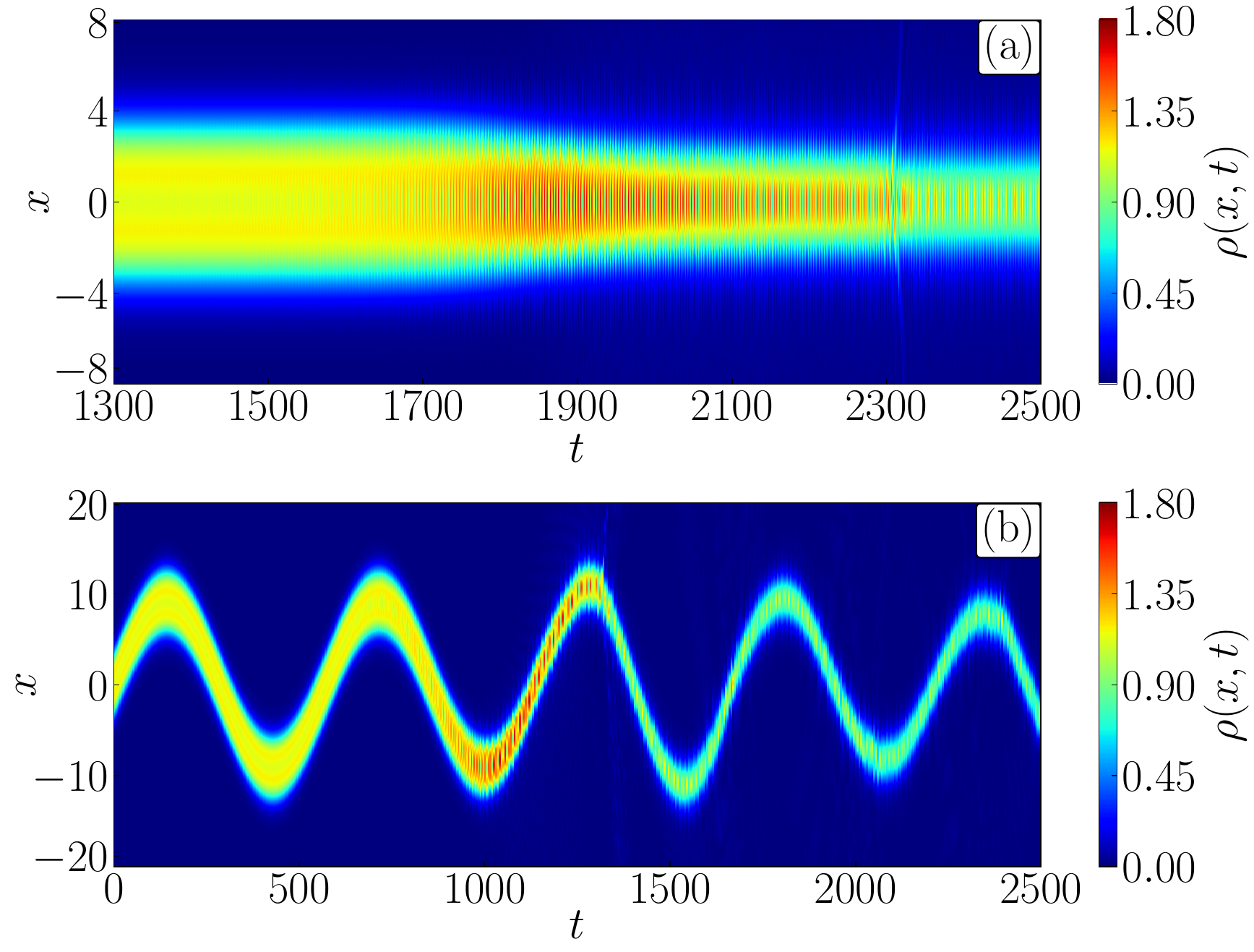}
		\caption{
		For simulations of ABS model~($m=-1$ and $\kappa=0$) (a) without and (b) with harmonic potential, $V(x)=(V_2/2) x^2$, with $V_2=10^{-4}$, time evolution of the charge density, $\rho(x,t)=|u(x,t)|^2+|v(x,t)|^2$.
		The numerical step is fixed at $h=10^{-3}$, the simulated length is $L=100$, and the initial condition is the ansatz  \eqref{eq26a}-\eqref{eq26b} with $q(0)=0$, $\dot{q}(0)=0.1$, $\omega=0.72$ and   
		$p(0)$ satisfying Eq.\ \eqref{eq:p}.}
	\label{fig:13} 	
\end{figure}

These instabilities at low frequencies persist even when the absorbing boundary conditions are added to the ABS  model, with and 
without potential. 
	In fact, charge and energy relative errors of order of $10^{-1}$ are found for frequencies in the interval $\omega \in [0.71,0.75]$ and harmonic potential with $V_2=10^{-4}$, even when the absorbing boundary conditions are applied. In the opposite case, errors lower than $10^{-5}$ are achieved for $\omega \in [0.76,0.99]$ and the same harmonic potential  without using absorbing boundary conditions. Furthermore, the 
simulations corroborate the numerical solutions of the equations of motion of the collective coordinates.

According to these results, neither Lakoba's method nor absorbing 
boundary conditions prevent soliton's  instability reported in Fig.~5 of Ref. \cite{mertens:2021} 
for the harmonic potential, and frequency $\omega=0.74$. 
Indeed, the soliton's profile is clearly distorted (see 
 \autoref{fig:abshar_w_71e-2_V2_1e-4}(a)),  
 and the errors of charge  grow for several sets of parameters (see \autoref{fig:abshar_w_71e-2_V2_1e-4}(b)). Moreover, 
the computed velocity, soliton position, and momentum from simulation deviate from the oscillatory character predicted by the CC theory.  Note that, for $\omega=0.74$ these signs of instability appear mostly in the simulations after  $t \approx 2500$ as in Fig.~5 of Ref. \cite{mertens:2021}, where the Runge-Kutta method was employed.  Similar behavior is observed when $\omega \in [0.71,0.775]$, and even when the time step is decreased up to $h=10^{-4}$.

\begin{figure}[h!]
	\includegraphics[width=0.99\linewidth]{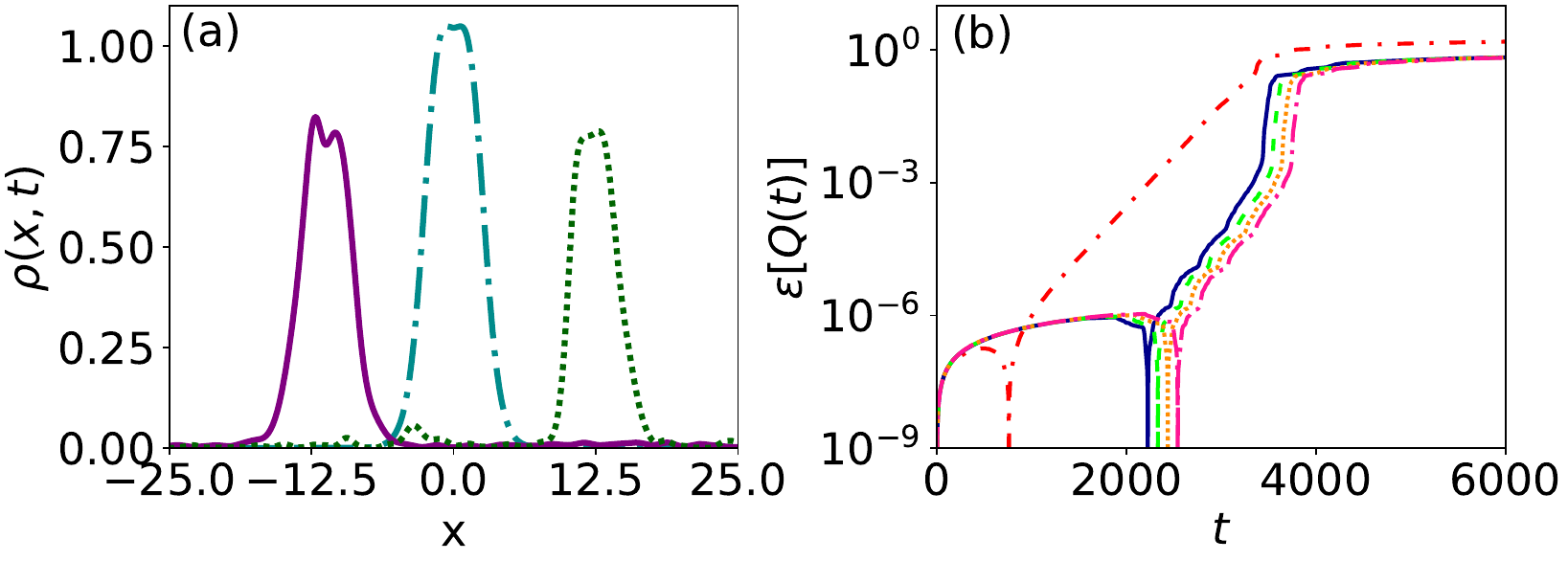}
	\caption{
	 Simulations of ABS model~($m=-1$ and $\kappa=0$) with harmonic potential, $V(x)=(V_2/2) x^2$, with $V_2=10^{-4}$. (a)
	Soliton charge density, $\rho(x,t)=|u(x,t)|^2+|v(x,t)|^2$, at 
		fixed times $t=0$~(dash-dotted cyan line), $t=3350$~(solid purple line), and $t=6000$~(dotted dark-green line) obtained from numerical simulations. 
		(b) Relative error of the charge for the simulated lengths $L=100$~(loosely dashed-dotted red line), $L=200$~(solid dark-blue line), $L=300$~(dashed green line), $L=400$~(dotted orange line) and $L=500$~(dashed-dotted magenta line).  
		The numerical step-size is fixed at $h=10^{-3}$, $x \in [-L,L]$, and the initial condition is the ansatz \eqref{eq26a}-\eqref{eq26b} with $q(0)=0$, $\dot{q}(0)=0.1$, $\omega=0.74$, 
		and $p(0)$ satisfying Eq.\ \eqref{eq:p}.}
	\label{fig:abshar_w_71e-2_V2_1e-4}
\end{figure}

\section{Conclusions} \label{sec6}

This work investigates the stability of the solitary waves in the Gross-Neveu and Alexeeva-Barashenkov-Saxena models, under linear and harmonic spatial potentials. The main difference between the two equations is the nonlinearity. Numerical simulations of these equations were performed by means of Lakoba's method, which, instead of carrying out  the  integration  in the original variables, employs two characteristic coordinates. The solution is computed by moving along constant characteristic lines. A second-order predictor-corrector method is 
 applied to obtain the numerical solution. This method was originally suggested by Lakoba  in Ref. \cite{lakoba:2018}, in order to integrate the Gross-Neveu equation without perturbations. Here, the algorithm is extended to the  Gross-Neveu and Alexeeva-Balashenkov-Saxena models under external potentials, where 
the characteristic variables explicitly appear
in the inhomogeneous terms. 

In order to study the stability of the numerical algorithm, the evolution of the charge and the energy 
are compared  with their initial values along the integration time. Both magnitudes are conserved when the potential is real and only space-dependent. Furthermore, in the case of the linear potential, the evolution equation for the field momentum is solved exactly regardless of the ansatz. This function is also employed to verify the validity of the numerical algorithm. All these comparisons allow  
the integration step $h$ to be properly chosen. The convergence of the numerical algorithm has been studied by decreasing the value of $h$ and by taking different  intervals of the simulated spatial-coordinate  $x \in [-L,L]$. The value of $h=10^{-3}$ already guarantees the convergence of our numerical simulations. 

The two spinor numerical solutions are used to compute the position and  the velocity of the center of mass, as well as the  charge density, the charge itself, the field energy, and the field momentum. The continuity equation for the charge is used to derive an expression for the velocity of the center of mass  and it is shown that the velocity is proportional to the integral of the  current density.  
The evolution of the charge density shows stable traveling solitary waves over long integration times for almost all considered parameters, in contrast with the results reported in Refs. \cite{mertens:2012, mertens:2021}. For the integration step $h=10^{-3}$, the relative errors of the energy and the charge are of the order of  $10^{-5}$ or less than this value. These errors can be reduced by decreasing the value of $h$, although the simulations become time consuming. The instabilities appeared  for low frequencies in the Gross-Neveu model  
are removed by considering absorbing boundary conditions, however, the soliton position
is 	modified by the absorbing boundary conditions. 
The unstable solitary waves for low frequencies in the ABS equation remain unstable when the harmonic potential is considered. 

The solitary wave solution is approximated by an ansatz with only two collective coordinates, namely the soliton position, $q(t)$, and momentum, $p(t)$. For the  unforced problem, the suggested ansatz 
becomes the exact moving solitary waves of the Gross-Neveu or Alexeeva-Balashenkov-Saxena equations. The evolution equations have been found by using the time variation of the field energy and the field momentum. 
The main advantage of this procedure is that the equations of motion always fulfill the continuity equations for the charge, momentum, and energy.
The equations of motion satisfied by the collective coordinates for the GN and ABS systems are the same, except for the range of the parameters $\omega$, and the values of the mass at rest $M_0$, and of the charge $Q$ (see Table \ref{tab1}).              

 To summarize, the solitary waves of Gross-Neveu and Alexeeva-Barashenkov-Saxena equations under harmonic  potentials are  numerically stables 
	over a long period of integration, and 
	within the range of parameters  
considered herein, except the ABS solitons for low-frequency limit. 
Therefore, almost all instabilities previously reported  are removed by means of Lakoba's method. It is worth noting that, although this numerical algorithm is used here for spatial potential, it can be straightforwardly extended when 
the external potential is also time-dependent. 
Despite the 
combination of this method with absorbing boundary conditions can remove almost all observed instabilities, it would be important to develop a numerical algorithm with exceptional discretizations that preserves both charge and energy.


\begin{acknowledgments}
	The authors thank Nora Alexeeva for providing us the data of numerical simulations of Fig. 1 of Ref. \cite{alexeeva:2019}.  
	This research was partially funded by the Spanish projects PID2020-113390GB-I00 (MICIN), PY20$\_$00082 
	(Junta de Andalucia)  and A-FQM-52-UGR20 (ERDF-University of Granada) and the Andalusian research group FQM-207.
\end{acknowledgments}

%


\begin{thebibliography}{30}%
\makeatletter
\providecommand \@ifxundefined [1]{%
 \@ifx{#1\undefined}
}%
\providecommand \@ifnum [1]{%
 \ifnum #1\expandafter \@firstoftwo
 \else \expandafter \@secondoftwo
 \fi
}%
\providecommand \@ifx [1]{%
 \ifx #1\expandafter \@firstoftwo
 \else \expandafter \@secondoftwo
 \fi
}%
\providecommand \natexlab [1]{#1}%
\providecommand \enquote  [1]{``#1''}%
\providecommand \bibnamefont  [1]{#1}%
\providecommand \bibfnamefont [1]{#1}%
\providecommand \citenamefont [1]{#1}%
\providecommand \href@noop [0]{\@secondoftwo}%
\providecommand \href [0]{\begingroup \@sanitize@url \@href}%
\providecommand \@href[1]{\@@startlink{#1}\@@href}%
\providecommand \@@href[1]{\endgroup#1\@@endlink}%
\providecommand \@sanitize@url [0]{\catcode `\\12\catcode `\$12\catcode
  `\&12\catcode `\#12\catcode `\^12\catcode `\_12\catcode `\%12\relax}%
\providecommand \@@startlink[1]{}%
\providecommand \@@endlink[0]{}%
\providecommand \url  [0]{\begingroup\@sanitize@url \@url }%
\providecommand \@url [1]{\endgroup\@href {#1}{\urlprefix }}%
\providecommand \urlprefix  [0]{URL }%
\providecommand \Eprint [0]{\href }%
\providecommand \doibase [0]{https://doi.org/}%
\providecommand \selectlanguage [0]{\@gobble}%
\providecommand \bibinfo  [0]{\@secondoftwo}%
\providecommand \bibfield  [0]{\@secondoftwo}%
\providecommand \translation [1]{[#1]}%
\providecommand \BibitemOpen [0]{}%
\providecommand \bibitemStop [0]{}%
\providecommand \bibitemNoStop [0]{.\EOS\space}%
\providecommand \EOS [0]{\spacefactor3000\relax}%
\providecommand \BibitemShut  [1]{\csname bibitem#1\endcsname}%
\let\auto@bib@innerbib\@empty
\bibitem [{\citenamefont {Parmentier}(1967)}]{parmentier:1967}%
  \BibitemOpen
  \bibfield  {author} {\bibinfo {author} {\bibfnamefont {R.~D.}\ \bibnamefont
  {Parmentier}},\ }\bibfield  {title} {\enquote {\bibinfo {title} {{Stability
  analysis of neuristor waveforms}},}\ }\href
  {https://doi.org/10.1109/PROC.1967.5865} {\bibfield  {journal} {\bibinfo
  {journal} {Proceedings of the IEEE}\ }\textbf {\bibinfo {volume} {55}},\
  \bibinfo {pages} {1498} (\bibinfo {year} {1967})}\BibitemShut {NoStop}%
\bibitem [{\citenamefont {Scott}(1969)}]{scott:1969}%
  \BibitemOpen
  \bibfield  {author} {\bibinfo {author} {\bibfnamefont {A.~C.}\ \bibnamefont
  {Scott}},\ }\bibfield  {title} {\enquote {\bibinfo {title} {{Waveform
  stability on a nonlinear Klein-Gordon equation}},}\ }\href@noop {} {\bibfield
   {journal} {\bibinfo  {journal} {Proceedings of the IEEE}\ }\textbf {\bibinfo
  {volume} {57}},\ \bibinfo {pages} {1338} (\bibinfo {year}
  {1969})}\BibitemShut {NoStop}%
\bibitem [{\citenamefont {Rubinstein}(1970)}]{rubinstein:1970}%
  \BibitemOpen
  \bibfield  {author} {\bibinfo {author} {\bibfnamefont {J.}~\bibnamefont
  {Rubinstein}},\ }\bibfield  {title} {\enquote {\bibinfo {title}
  {{Sine‐Gordon Equation}},}\ }\href {https://doi.org/10.1063/1.1665057}
  {\bibfield  {journal} {\bibinfo  {journal} {{J. Math. Phys.}}\ }\textbf
  {\bibinfo {volume} {11}},\ \bibinfo {pages} {258} (\bibinfo {year}
  {1970})}\BibitemShut {NoStop}%
\bibitem [{\citenamefont {Zakharov}(1967)}]{zakharov:1967}%
  \BibitemOpen
  \bibfield  {author} {\bibinfo {author} {\bibfnamefont {V.}~\bibnamefont
  {Zakharov}},\ }\bibfield  {title} {\enquote {\bibinfo {title} {{Instability
  of self-focusing of light}},}\ }\href@noop {} {\bibfield  {journal} {\bibinfo
   {journal} {Zh. \'Eksp. Teor. Fiz}\ }\textbf {\bibinfo {volume} {53}},\
  \bibinfo {pages} {1735} (\bibinfo {year} {1967})}\BibitemShut {NoStop}%
\bibitem [{\citenamefont {Vakhitov}\ and\ \citenamefont
  {Kolokolov}(1973)}]{vakhitov:1973}%
  \BibitemOpen
  \bibfield  {author} {\bibinfo {author} {\bibfnamefont {N.~G.}\ \bibnamefont
  {Vakhitov}}\ and\ \bibinfo {author} {\bibfnamefont {A.~A.}\ \bibnamefont
  {Kolokolov}},\ }\bibfield  {title} {\enquote {\bibinfo {title} {Stationary
  solutions of the wave equation in the medium with nonlinearity saturation},}\
  }\href@noop {} {\bibfield  {journal} {\bibinfo  {journal} {Radiophys. Quantum
  Electron.}\ }\textbf {\bibinfo {volume} {16}},\ \bibinfo {pages} {783}
  (\bibinfo {year} {1973})}\BibitemShut {NoStop}%
\bibitem [{\citenamefont {Chugunova}(2007)}]{chugunova:2007}%
  \BibitemOpen
  \bibfield  {author} {\bibinfo {author} {\bibfnamefont {M.}~\bibnamefont
  {Chugunova}},\ }\href@noop {} {\emph {\bibinfo {title} {{Spectral stability
  of nonlinear waves in dynamical systems}}}}\ (\bibinfo  {publisher} {Doctoral
  Thesis},\ \bibinfo {address} {McMaster University, Hamilton, Ontario,
  Canada},\ \bibinfo {year} {2007})\BibitemShut {NoStop}%
\bibitem [{\citenamefont {Berkolaiko}\ and\ \citenamefont
  {Comech}(2012)}]{berkolaiko:2012}%
  \BibitemOpen
  \bibfield  {author} {\bibinfo {author} {\bibfnamefont {G.}~\bibnamefont
  {Berkolaiko}}\ and\ \bibinfo {author} {\bibfnamefont {A.}~\bibnamefont
  {Comech}},\ }\bibfield  {title} {\enquote {\bibinfo {title} {{On Spectral
  Stability of Solitary Waves of Nonlinear Dirac Equation in 1D}},}\ }\href
  {https://doi.org/10.1051/mmnp/20127202} {\bibfield  {journal} {\bibinfo
  {journal} {Math. Model. Nat. Phenom.}\ }\textbf {\bibinfo {volume} {7}},\
  \bibinfo {pages} {13} (\bibinfo {year} {2012})}\BibitemShut {NoStop}%
\bibitem [{\citenamefont {Bogolubsky}(1979)}]{bogolubsky:1979}%
  \BibitemOpen
  \bibfield  {author} {\bibinfo {author} {\bibfnamefont {I.~L.}\ \bibnamefont
  {Bogolubsky}},\ }\bibfield  {title} {\enquote {\bibinfo {title} {{On spinor
  soliton stability}},}\ }\href@noop {} {\bibfield  {journal} {\bibinfo
  {journal} {Phys. Lett. A}\ }\textbf {\bibinfo {volume} {73}},\ \bibinfo
  {pages} {87} (\bibinfo {year} {1979})}\BibitemShut {NoStop}%
\bibitem [{\citenamefont {Shao}\ \emph {et~al.}(2014)\citenamefont {Shao},
  \citenamefont {Quintero}, \citenamefont {Mertens}, \citenamefont {Cooper},
  \citenamefont {Khare},\ and\ \citenamefont {Saxena}}]{shao:2014}%
  \BibitemOpen
  \bibfield  {author} {\bibinfo {author} {\bibfnamefont {S.}~\bibnamefont
  {Shao}}, \bibinfo {author} {\bibfnamefont {N.~R.}\ \bibnamefont {Quintero}},
  \bibinfo {author} {\bibfnamefont {F.~G.}\ \bibnamefont {Mertens}}, \bibinfo
  {author} {\bibfnamefont {F.}~\bibnamefont {Cooper}}, \bibinfo {author}
  {\bibfnamefont {A.}~\bibnamefont {Khare}},\ and\ \bibinfo {author}
  {\bibfnamefont {A.}~\bibnamefont {Saxena}},\ }\bibfield  {title} {\enquote
  {\bibinfo {title} {{Stability of solitary waves in the nonlinear Dirac
  equation with arbitrary nonlinearity}},}\ }\href@noop {} {\bibfield
  {journal} {\bibinfo  {journal} {Phys. Rev. E}\ }\textbf {\bibinfo {volume}
  {90}},\ \bibinfo {pages} {032915} (\bibinfo {year} {2014})}\BibitemShut
  {NoStop}%
\bibitem [{\citenamefont {Xu}, \citenamefont {Shao},\ and\ \citenamefont
  {Tang}(2013)}]{xu:2013}%
  \BibitemOpen
  \bibfield  {author} {\bibinfo {author} {\bibfnamefont {J.}~\bibnamefont
  {Xu}}, \bibinfo {author} {\bibfnamefont {S.}~\bibnamefont {Shao}},\ and\
  \bibinfo {author} {\bibfnamefont {H.}~\bibnamefont {Tang}},\ }\bibfield
  {title} {\enquote {\bibinfo {title} {{Numerical methods for nonlinear Dirac
  equation}},}\ }\href
  {https://doi.org/https://doi.org/10.1016/j.jcp.2013.03.031} {\bibfield
  {journal} {\bibinfo  {journal} {J. Comput. Phys.}\ }\textbf {\bibinfo
  {volume} {245}},\ \bibinfo {pages} {131} (\bibinfo {year}
  {2013})}\BibitemShut {NoStop}%
\bibitem [{\citenamefont {Lakoba}(2018)}]{lakoba:2018}%
  \BibitemOpen
  \bibfield  {author} {\bibinfo {author} {\bibfnamefont {T.}~\bibnamefont
  {Lakoba}},\ }\bibfield  {title} {\enquote {\bibinfo {title} {{Numerical study
  of solitary wave stability in cubic nonlinear Dirac equations in 1D}},}\
  }\href@noop {} {\bibfield  {journal} {\bibinfo  {journal} {Phys. Lett. A}\
  }\textbf {\bibinfo {volume} {382}},\ \bibinfo {pages} {300} (\bibinfo {year}
  {2018})}\BibitemShut {NoStop}%
\bibitem [{\citenamefont {Lakoba}\ and\ \citenamefont
  {Jewell}(2021)}]{lakoba:2021}%
  \BibitemOpen
  \bibfield  {author} {\bibinfo {author} {\bibfnamefont {T.~I.}\ \bibnamefont
  {Lakoba}}\ and\ \bibinfo {author} {\bibfnamefont {J.~S.}\ \bibnamefont
  {Jewell}},\ }\bibfield  {title} {\enquote {\bibinfo {title} {{Higher-order
  explicit schemes based on the method of characteristics for hyperbolic
  equations with crossing straight-line characteristics}},}\ }\href@noop {}
  {\bibfield  {journal} {\bibinfo  {journal} {Numer. Methods Partial Differ.
  Equ.}\ }\textbf {\bibinfo {volume} {37}},\ \bibinfo {pages} {2742} (\bibinfo
  {year} {2021})}\BibitemShut {NoStop}%
\bibitem [{\citenamefont {Cuevas-Maraver}\ \emph {et~al.}(2015)\citenamefont
  {Cuevas-Maraver}, \citenamefont {Kevrekidis}, \citenamefont {Saxena},
  \citenamefont {Cooper},\ and\ \citenamefont {Mertens}}]{cuevas:2015}%
  \BibitemOpen
  \bibfield  {author} {\bibinfo {author} {\bibfnamefont {J.}~\bibnamefont
  {Cuevas-Maraver}}, \bibinfo {author} {\bibfnamefont {P.}~\bibnamefont
  {Kevrekidis}}, \bibinfo {author} {\bibfnamefont {A.}~\bibnamefont {Saxena}},
  \bibinfo {author} {\bibfnamefont {F.}~\bibnamefont {Cooper}},\ and\ \bibinfo
  {author} {\bibfnamefont {F.}~\bibnamefont {Mertens}},\ }\href@noop {} {\emph
  {\bibinfo {title} {{Ordinary and Partial Differential Equations}}}}\
  (\bibinfo  {publisher} {Nova Sciences},\ \bibinfo {address} {New York},\
  \bibinfo {year} {2015})\ Chap.\ \bibinfo {chapter} {{Solitary Waves in the
  Nonlinear Dirac Equation at the Continuum Limit: Stability and
  Dynamics}}\BibitemShut {NoStop}%
\bibitem [{\citenamefont {Bogdan}, \citenamefont {Kosevich},\ and\
  \citenamefont {Manzhos}(1985)}]{bogdan:1985}%
  \BibitemOpen
  \bibfield  {author} {\bibinfo {author} {\bibfnamefont {M.~M.}\ \bibnamefont
  {Bogdan}}, \bibinfo {author} {\bibfnamefont {A.~M.}\ \bibnamefont
  {Kosevich}},\ and\ \bibinfo {author} {\bibfnamefont {I.~V.}\ \bibnamefont
  {Manzhos}},\ }\bibfield  {title} {\enquote {\bibinfo {title} {{Stabilization
  of a magnetic soliton (bion) as a result of parametric excitation of
  one-dimensional ferromagnet}},}\ }\href@noop {} {\bibfield  {journal}
  {\bibinfo  {journal} {{Sov. J. Low Temp. Phys.}}\ }\textbf {\bibinfo {volume}
  {11}},\ \bibinfo {pages} {547} (\bibinfo {year} {1985})}\BibitemShut
  {NoStop}%
\bibitem [{\citenamefont {Quintero}\ and\ \citenamefont
  {Sánchez-Rey}(2019)}]{quintero:2019b}%
  \BibitemOpen
  \bibfield  {author} {\bibinfo {author} {\bibfnamefont {N.~R.}\ \bibnamefont
  {Quintero}}\ and\ \bibinfo {author} {\bibfnamefont {B.}~\bibnamefont
  {Sánchez-Rey}},\ }\bibfield  {title} {\enquote {\bibinfo {title} {{Exact
  stationary solutions of the parametrically driven and damped nonlinear Dirac
  equation}},}\ }\href@noop {} {\bibfield  {journal} {\bibinfo  {journal}
  {Chaos}\ }\textbf {\bibinfo {volume} {29}},\ \bibinfo {pages} {093129}
  (\bibinfo {year} {2019})}\BibitemShut {NoStop}%
\bibitem [{\citenamefont {Nogami}, \citenamefont {Toyama},\ and\ \citenamefont
  {Zhao}(1995)}]{nogami:1995}%
  \BibitemOpen
  \bibfield  {author} {\bibinfo {author} {\bibfnamefont {Y.}~\bibnamefont
  {Nogami}}, \bibinfo {author} {\bibfnamefont {F.~M.}\ \bibnamefont {Toyama}},\
  and\ \bibinfo {author} {\bibfnamefont {Z.}~\bibnamefont {Zhao}},\ }\bibfield
  {title} {\enquote {\bibinfo {title} {{Nonlinear Dirac soliton in an external
  field}},}\ }\href {http://stacks.iop.org/0305-4470/28/i=5/a=025} {\bibfield
  {journal} {\bibinfo  {journal} {J. Phys. A: Math. Gen.}\ }\textbf {\bibinfo
  {volume} {28}},\ \bibinfo {pages} {1413} (\bibinfo {year}
  {1995})}\BibitemShut {NoStop}%
\bibitem [{\citenamefont {Mertens}\ \emph {et~al.}(2012)\citenamefont
  {Mertens}, \citenamefont {Quintero}, \citenamefont {Cooper}, \citenamefont
  {Khare},\ and\ \citenamefont {Saxena}}]{mertens:2012}%
  \BibitemOpen
  \bibfield  {author} {\bibinfo {author} {\bibfnamefont {F.~G.}\ \bibnamefont
  {Mertens}}, \bibinfo {author} {\bibfnamefont {N.~R.}\ \bibnamefont
  {Quintero}}, \bibinfo {author} {\bibfnamefont {F.}~\bibnamefont {Cooper}},
  \bibinfo {author} {\bibfnamefont {A.}~\bibnamefont {Khare}},\ and\ \bibinfo
  {author} {\bibfnamefont {A.}~\bibnamefont {Saxena}},\ }\bibfield  {title}
  {\enquote {\bibinfo {title} {{Nonlinear Dirac equation solitary waves in
  external fields}},}\ }\href@noop {} {\bibfield  {journal} {\bibinfo
  {journal} {Phys. Rev. E}\ }\textbf {\bibinfo {volume} {86}},\ \bibinfo
  {pages} {046602} (\bibinfo {year} {2012})}\BibitemShut {NoStop}%
\bibitem [{\citenamefont {Lakoba}(2020)}]{lakoba:2020}%
  \BibitemOpen
  \bibfield  {author} {\bibinfo {author} {\bibfnamefont {T.}~\bibnamefont
  {Lakoba}},\ }\bibfield  {title} {\enquote {\bibinfo {title} {{Study of
  instability of the Fourier split-step method for the massive Gross-Neveu
  model}},}\ }\href@noop {} {\bibfield  {journal} {\bibinfo  {journal} {J.
  Comput. Phys.}\ }\textbf {\bibinfo {volume} {402}},\ \bibinfo {pages}
  {109100} (\bibinfo {year} {2020})}\BibitemShut {NoStop}%
\bibitem [{\citenamefont {Mertens}, \citenamefont {Sánchez-Rey},\ and\
  \citenamefont {Quintero}(2021)}]{mertens:2021}%
  \BibitemOpen
  \bibfield  {author} {\bibinfo {author} {\bibfnamefont {F.~G.}\ \bibnamefont
  {Mertens}}, \bibinfo {author} {\bibfnamefont {B.}~\bibnamefont
  {Sánchez-Rey}},\ and\ \bibinfo {author} {\bibfnamefont {N.~R.}\ \bibnamefont
  {Quintero}},\ }\bibfield  {title} {\enquote {\bibinfo {title} {{Soliton
  dynamics in the ABS nonlinear spinor model with external fields}},}\
  }\href@noop {} {\bibfield  {journal} {\bibinfo  {journal} {J. of Phys A:
  Math. and Theo.}\ }\textbf {\bibinfo {volume} {54}},\ \bibinfo {pages}
  {405702} (\bibinfo {year} {2021})}\BibitemShut {NoStop}%
\bibitem [{\citenamefont {Alexeeva}, \citenamefont {Barashenkov},\ and\
  \citenamefont {Saxena}(2019)}]{alexeeva:2019}%
  \BibitemOpen
  \bibfield  {author} {\bibinfo {author} {\bibfnamefont {N.~V.}\ \bibnamefont
  {Alexeeva}}, \bibinfo {author} {\bibfnamefont {I.~V.}\ \bibnamefont
  {Barashenkov}},\ and\ \bibinfo {author} {\bibfnamefont {A.}~\bibnamefont
  {Saxena}},\ }\bibfield  {title} {\enquote {\bibinfo {title} {{Spinor solitons
  and their PT-symmetric offspring}},}\ }\href@noop {} {\bibfield  {journal}
  {\bibinfo  {journal} {Ann. Phys.}\ }\textbf {\bibinfo {volume} {403}},\
  \bibinfo {pages} {198} (\bibinfo {year} {2019})}\BibitemShut {NoStop}%
\bibitem [{\citenamefont {McLaughlin}\ and\ \citenamefont
  {Scott}(1978)}]{mclaughlin:1978}%
  \BibitemOpen
  \bibfield  {author} {\bibinfo {author} {\bibfnamefont {D.~W.}\ \bibnamefont
  {McLaughlin}}\ and\ \bibinfo {author} {\bibfnamefont {A.~C.}\ \bibnamefont
  {Scott}},\ }\bibfield  {title} {\enquote {\bibinfo {title} {{Perturbation
  analysis of fluxon dynamics}},}\ }\href@noop {} {\bibfield  {journal}
  {\bibinfo  {journal} {Phys. Rev. A}\ }\textbf {\bibinfo {volume} {18}},\
  \bibinfo {pages} {1652} (\bibinfo {year} {1978})}\BibitemShut {NoStop}%
\bibitem [{\citenamefont {Karpman}(1979)}]{karpman:1979}%
  \BibitemOpen
  \bibfield  {author} {\bibinfo {author} {\bibfnamefont {V.~I.}\ \bibnamefont
  {Karpman}},\ }\bibfield  {title} {\enquote {\bibinfo {title} {{Soliton
  Evolution in the Presence of Perturbation}},}\ }\href@noop {} {\bibfield
  {journal} {\bibinfo  {journal} {Phys. Scripta}\ }\textbf {\bibinfo {volume}
  {20}},\ \bibinfo {pages} {462} (\bibinfo {year} {1979})}\BibitemShut
  {NoStop}%
\bibitem [{\citenamefont {Maimistov}(1993)}]{maimistov:1993}%
  \BibitemOpen
  \bibfield  {author} {\bibinfo {author} {\bibfnamefont {A.~I.}\ \bibnamefont
  {Maimistov}},\ }\bibfield  {title} {\enquote {\bibinfo {title} {{Evolution of
  solitary waves which are approximately solitons of a nonlinear Schr\"odinger
  equation}},}\ }\href@noop {} {\bibfield  {journal} {\bibinfo  {journal} {Zh.
  Eksp. Teor. Fiz.}\ }\textbf {\bibinfo {volume} {104}},\ \bibinfo {pages}
  {3620} (\bibinfo {year} {1993})}\BibitemShut {NoStop}%
\bibitem [{\citenamefont {Quintero}, \citenamefont {Mertens},\ and\
  \citenamefont {Bishop}(2010)}]{quintero:2010}%
  \BibitemOpen
  \bibfield  {author} {\bibinfo {author} {\bibfnamefont {N.~R.}\ \bibnamefont
  {Quintero}}, \bibinfo {author} {\bibfnamefont {F.~G.}\ \bibnamefont
  {Mertens}},\ and\ \bibinfo {author} {\bibfnamefont {A.~R.}\ \bibnamefont
  {Bishop}},\ }\bibfield  {title} {\enquote {\bibinfo {title} {{Generalized
  traveling-wave method, variational approach, and modified conserved
  quantities for the perturbed nonlinear Schr\"odinger equation}},}\ }\href
  {https://doi.org/10.1103/PhysRevE.82.016606} {\bibfield  {journal} {\bibinfo
  {journal} {Phys. Rev. E}\ }\textbf {\bibinfo {volume} {82}},\ \bibinfo
  {pages} {016606} (\bibinfo {year} {2010})}\BibitemShut {NoStop}%
\bibitem [{\citenamefont {Alvarez}\ and\ \citenamefont
  {Carreras}(1981)}]{alvarez:1981}%
  \BibitemOpen
  \bibfield  {author} {\bibinfo {author} {\bibfnamefont {A.}~\bibnamefont
  {Alvarez}}\ and\ \bibinfo {author} {\bibfnamefont {B.}~\bibnamefont
  {Carreras}},\ }\bibfield  {title} {\enquote {\bibinfo {title} {{Interaction
  dynamics for the solitary waves of a nonlinear Dirac model}},}\ }\href@noop
  {} {\bibfield  {journal} {\bibinfo  {journal} {Phys. Lett. A}\ }\textbf
  {\bibinfo {volume} {86}},\ \bibinfo {pages} {327} (\bibinfo {year}
  {1981})}\BibitemShut {NoStop}%
\bibitem [{\citenamefont {Salerno}\ and\ \citenamefont
  {Zolotaryuk}(2002)}]{salerno:2002}%
  \BibitemOpen
  \bibfield  {author} {\bibinfo {author} {\bibfnamefont {M.}~\bibnamefont
  {Salerno}}\ and\ \bibinfo {author} {\bibfnamefont {Y.}~\bibnamefont
  {Zolotaryuk}},\ }\bibfield  {title} {\enquote {\bibinfo {title} {{Soliton
  ratchetlike dynamics by ac forces with harmonic mixing}},}\ }\href@noop {}
  {\bibfield  {journal} {\bibinfo  {journal} {Phys. Rev. E}\ }\textbf {\bibinfo
  {volume} {65}},\ \bibinfo {pages} {056603} (\bibinfo {year}
  {2002})}\BibitemShut {NoStop}%
\bibitem [{\citenamefont {Quintero}, \citenamefont {S\'anchez},\ and\
  \citenamefont {Mertens}(2001)}]{quintero:2001}%
  \BibitemOpen
  \bibfield  {author} {\bibinfo {author} {\bibfnamefont {N.~R.}\ \bibnamefont
  {Quintero}}, \bibinfo {author} {\bibfnamefont {A.}~\bibnamefont
  {S\'anchez}},\ and\ \bibinfo {author} {\bibfnamefont {F.~G.}\ \bibnamefont
  {Mertens}},\ }\bibfield  {title} {\enquote {\bibinfo {title} {{Anomalies of
  ac driven solitary waves with internal modes: Nonparametric resonances
  induced by parametric forces}},}\ }\href
  {https://doi.org/10.1103/PhysRevE.64.046601} {\bibfield  {journal} {\bibinfo
  {journal} {Phys. Rev. E}\ }\textbf {\bibinfo {volume} {64}},\ \bibinfo
  {pages} {046601} (\bibinfo {year} {2001})}\BibitemShut {NoStop}%
\bibitem [{\citenamefont {Morales-Molina}\ \emph {et~al.}(2006)\citenamefont
  {Morales-Molina}, \citenamefont {Quintero}, \citenamefont {S\'{a}nchez},\
  and\ \citenamefont {Mertens}}]{morales-molina:2006}%
  \BibitemOpen
  \bibfield  {author} {\bibinfo {author} {\bibfnamefont {L.}~\bibnamefont
  {Morales-Molina}}, \bibinfo {author} {\bibfnamefont {N.~R.}\ \bibnamefont
  {Quintero}}, \bibinfo {author} {\bibfnamefont {A.}~\bibnamefont
  {S\'{a}nchez}},\ and\ \bibinfo {author} {\bibfnamefont {F.~G.}\ \bibnamefont
  {Mertens}},\ }\bibfield  {title} {\enquote {\bibinfo {title} {{Soliton
  ratchets in homogeneous nonlinear Klein-Gordon systems}},}\ }\href@noop {}
  {\bibfield  {journal} {\bibinfo  {journal} {Chaos}\ }\textbf {\bibinfo
  {volume} {16}},\ \bibinfo {pages} {013117} (\bibinfo {year}
  {2006})}\BibitemShut {NoStop}%
\bibitem [{\citenamefont {Lee}, \citenamefont {Kuo},\ and\ \citenamefont
  {Gavrielides}(1975)}]{lee:1975}%
  \BibitemOpen
  \bibfield  {author} {\bibinfo {author} {\bibfnamefont {S.~Y.}\ \bibnamefont
  {Lee}}, \bibinfo {author} {\bibfnamefont {T.~K.}\ \bibnamefont {Kuo}},\ and\
  \bibinfo {author} {\bibfnamefont {A.}~\bibnamefont {Gavrielides}},\
  }\bibfield  {title} {\enquote {\bibinfo {title} {{Exact localized solutions
  of two-dimensional field theories of massive fermions with Fermi
  interactions}},}\ }\href@noop {} {\bibfield  {journal} {\bibinfo  {journal}
  {Phys. Rev. D}\ }\textbf {\bibinfo {volume} {12}},\ \bibinfo {pages} {2249}
  (\bibinfo {year} {1975})}\BibitemShut {NoStop}%
\bibitem [{\citenamefont {Birkhoff}\ and\ \citenamefont
  {Rota}(1989)}]{birkhoff:1989}%
  \BibitemOpen
  \bibfield  {author} {\bibinfo {author} {\bibfnamefont {G.}~\bibnamefont
  {Birkhoff}}\ and\ \bibinfo {author} {\bibfnamefont {G.~C.}\ \bibnamefont
  {Rota}},\ }\href@noop {} {\emph {\bibinfo {title} {{Ordinary differential
  equations}}}}\ (\bibinfo  {publisher} {Wiley},\ \bibinfo {year}
  {1989})\BibitemShut {NoStop}%
\end{thebibliography}
\end{document}